\documentclass[iop, apjl, twocolumn]{aastex631}
\usepackage{graphicx}
\usepackage{bm}
\usepackage{amsmath,amssymb}
\usepackage{booktabs}
\usepackage{color}
\usepackage{units}
\usepackage[normalem]{ulem}

\usepackage{booktabs}
\usepackage{soul}
\usepackage{tikz}
\usetikzlibrary{positioning, fit, arrows.meta}
\usepackage{bbding}
\usepackage{pifont}

\begin{document}

\title{How general relativity stacks up: \\
Testing gravity by coherently combining gravitational-wave observations}

\newcommand{\SPA}{School of Physics and Astronomy, Monash University, Clayton VIC 3800, Australia}
\newcommand{\OzGravMonash}{OzGrav: The ARC Centre of Excellence for Gravitational Wave Discovery, Clayton VIC 3800, Australia}
\newcommand{\UBC}{Department of Physics and Astronomy, University of British Columbia, Vancouver, British Columbia, V6T1Z4, Canada}

\author{Lachlan Passenger}
\email{lachlan.passenger1@monash.edu}
\affiliation{\SPA}
\affiliation{\OzGravMonash}

\author{Nir Guttman}
\affiliation{\SPA}
\affiliation{\OzGravMonash}

\author{Paul D. Lasky}
\affiliation{\SPA}
\affiliation{\OzGravMonash}
\affiliation{\UBC}

\author{Eric Thrane}
\affiliation{\SPA}
\affiliation{\OzGravMonash}

\date{\today}

\begin{abstract}

A core mission of gravitational-wave astronomy is to test general relativity (GR) in the high-curvature, high-velocity regime.
The field has already seen significant progress in this space, though a remaining challenge is how to combine data from a population of binary mergers. This can both boost the significance of potential GR violations, and develop a phenomenological picture of how deviations manifest in different mergers.
We propose a framework to coherently combine gravitational-wave measurements to build a population-level test of physics beyond GR.
We assume that deviations from GR are (i) strongest near the merger, (ii) vary smoothly with the total mass of the binary, and (iii) are most prominent in the $\ell=|m|=2$ mode.
While these assumptions do not describe every possible deviation of GR, they provide us with a plausible starting point upon which a more general framework can be built.
We apply our framework to 28 high signal-to-noise ratio binary black hole mergers from LIGO-Virgo-KAGRA's Fifth Gravitational-Wave Transient Catalog, finding no evidence for deviations from GR. Our framework allows one to show how deviations could vary as a function of total mass of the system. We argue that such a reconstruction can be used to convincingly distinguish between systematic waveform errors and genuine GR violations.
\end{abstract}

\section{Introduction}
Testing general relativity (GR) is one of the key science goals of the LIGO-Virgo-KAGRA collaboration \citep[LVK;][]{theligoscientificcollaborationAdvancedLIGO2015, acerneseAdvancedVirgo2nd2015a, akutsu_overview_2021} and gravitational-wave astronomy as a whole \citep[e.g.,][]{reitze_cosmic_2019, bailes_gravitational-wave_2021}. In particular, gravitational waves from compact binary mergers provide an opportunity to test GR in the strong-field, highly dynamical regime, where nonlinear dynamics are encoded directly in the gravitational-wave signal. 

As an example of strong-field tests, the no-hair theorem of GR predicts that the quasinormal mode spectrum of a black hole is determined uniquely by its mass and spin. This has been confirmed by evidence for the fundamental $(\ell=|m|=2)$ mode in GW150914 \citep[][]{collaboration_tests_2016}, as well as its first overtone (e.g., \citet[][]{isi_testing_2019}, though see \citet[][]{bustillo_black-hole_2021, cotesta_analysis_2022, chandra_black-hole_2025} for a different assessment). Evidence for the first overtone has also been found in GW231028 \citep[][]{wang_decisive_2025} and GW250114 \citep[][]{abac_black_2026}, while higher-order ringdown modes have been reported in GW190521 \citep[][]{capano_multimode_2023, siegel_ringdown_2023}, GW231123 \citep[][]{siegel_gw231123_2025, wang_detection_2025} and GW250114 \citep[][]{abac_black_2026, yang_contribution_2025}. Recently, GW250114 was used to confirm Hawking's area law \citep[][]{abac_gw250114_2025, chung_measuring_2026}. Other tests such as inspiral-merger-ringdown consistency checks \citep[e.g.,][]{hughesGoldenBinaryGravitationalWave2005, ghoshTestingGeneralRelativity2017, collaboration_tests_2016, testinggrgwtc1, collaboration_tests_2021, the_ligo_scientific_collaboration_tests_2025, collaboration_gwtc-40_2026-1}, tests of gravitational-wave generation using a parameterised post-newtonian framework \citep[e.g.,][]{collaboration_tests_2016, testinggrgwtc1, collaboration_tests_2021, the_ligo_scientific_collaboration_tests_2025, collaboration_gwtc-40_2026, testinggrgwtc5}, residual tests and excess power searches \citep[e.g.,][]{ collaboration_tests_2016, testinggrgwtc1, collaboration_tests_2021, edelmanConstrainingUnmodeledPhysics2021, the_ligo_scientific_collaboration_tests_2025, collaboration_gwtc-40_2026-1, testinggrgwtc5}, extra polarisation tests \citep[e.g.,][]{collaboration_tests_2021, wong_null-stream-based_2021, the_ligo_scientific_collaboration_tests_2025, cheung_search_2025, collaboration_gwtc-40_2026-1, testinggrgwtc5} and modified gravitational-wave dispersion \citep[e.g.,][]{testinggrgwtc1, collaboration_tests_2021, the_ligo_scientific_collaboration_tests_2025, collaboration_gwtc-40_2026} have found consistency between current gravitational-wave observations and GR.

While analysis of individual events provides valuable insights into physics beyond GR, it has significant limitations. First, the resolving power is primarily driven by the signal-to-noise ratio (SNR) of the loudest events.
Second, an event-by-event approach does not, by itself, determine \textit{how} deviations vary across different gravitational-wave events.
The obvious solution is to combine the information from multiple events, but the question of how best to do this turns out to be surprisingly difficult, subtle, and interesting.

The approach taken by \cite{maselli_parametrized_2020}---and further developed by \cite{carullo_enhancing_2021,maselli_black_2024}---is to assume that new physics is described by deviation parameters.
Specifically, they introduce spin-expansion coefficients that modify the usual relation between the dimensionless ringdown frequency and the remnant spin.
These coefficients are assumed to scale with the remnant mass according to a power law, with a spectral index determined by the specific choice of model.
One potential disadvantage of this approach is that it may not be sufficiently general to describe physics beyond GR.
That is, it is not obvious that physics beyond GR is well described by \cite{maselli_black_2024}'s \textsc{ParSpec} parameterization in which all deviation coefficients share the same mass scaling.

\citet{isi_hierarchical_2019}---further developed by \cite{zhong_multidimensional_2024}, and applied to various gravitational-wave transient catalogs by the LVK collaboration~\citep{collaboration_tests_2021,the_ligo_scientific_collaboration_tests_2025,collaboration_gwtc-40_2026}---provides a more general framework.
They too employ ringdown deviation parameters, e.g., describing shifts in ringdown frequencies and damping times.
Their model assumes that the deviation parameters for different events are not the same, but that they are drawn randomly from some distribution.
A related, more conservative approach by \citet[][]{chandra_testing_2026} models the catalog as a mixture in which some fraction of events obey GR exactly, while the remainder deviate from GR with no population constraint.
While these methods are extremely general, they do not exploit the physically reasonable expectation that deviations from GR are deterministic in nature rather than stochastic.
If we observe two identical binaries, we expect to see the same deviation from GR, but these frameworks do not impose such a condition.
As such, they do not provide a picture for \textit{how} different deviations change across different events\footnote{Our concern is partly addressed in \cite{payne_curvature_2024}, who extend the model of \citet{isi_hierarchical_2019} so that the mean of the distribution of deviation parameters varies with total mass. However, the model in \cite{payne_curvature_2024} is still fundamentally stochastic.}.

\cite{dideron_new_2023,dideron_agnostically_2026} gets rid of deviation parameters entirely, instead fitting strain deviations, which they allow to vary with the total mass of the binary.
This phenomenological approach allows for a wide range of deviations and, unlike the frameworks discussed above, does not restrict deviations to the ringdown part of the signal.
This seems well motivated because it is plausible that deviations from GR are most pronounced during the merger \textit{before} the ringdown.\footnote{We acknowledge the usefulness of ringdown tests. Unlike the merger, the ringdown can be described analytically, and different parts of the ringdown can be attributed to different physical processes in the perturbed black hole. This does not change our belief that deviations from GR are likely to be more prominent during the merger as the system is in a higher dynamic state.}
However, the \cite{dideron_agnostically_2026} model is still fundamentally stochastic such that identical binaries are not forced to exhibit identical deviations from GR.

In this paper, we introduce a framework for coherently combining gravitational-wave measurements by modeling the ``deviated strain''---the non-GR contribution to the gravitational-wave strain.
We aim to retain the determinism of strongly parameterised approaches like \cite{maselli_black_2024}, but with the flexibility of stochastic models like \cite{isi_hierarchical_2019}.
Following \cite{dideron_agnostically_2026}, we eschew deviation parameters in favor of a phenomenological model for the beyond-GR strain. 
This better reflects our prior beliefs about the morphology of deviations from GR in individual events, which we describe in more detail below.
Also, like \cite{dideron_agnostically_2026}, we do not restrict our attention to the ringdown, but consider the whole signal with a particular focus on dynamics through the merger phase where we suspect a deviation would be most apparent.

To this end, we employ the Gaussian-process model from \cite{passenger_gaussian_2025} to enforce our prior belief that the deviated strain is likely to be localized in time near the merger, oscillating at some characteristic frequency.
We additionally assume that deviated strain is determined entirely by the binary total mass, so that two events with the same total mass yield identical deviated strain.
However, two events with very different total mass may exhibit completely different deviated strain.
Finally, in order to account for each binary having different viewing angles, we hypothesize that the deviated strain is most pronounced in the $\ell=|m|=2$ mode.
We see this as a plausible starting point in a program of research to develop a fully general framework.

Using our framework, we construct a ``map'' showing how deviated strain varies with total mass.
This map provides a predictive model that can be tested against future measurements.
By studying the mass-dependence, we can determine if the deviated strain is consistent with waveform systematics---which are scale-invariant\footnote{In other words, the shape of a gravitational waveform described by GR does not depend on the total mass of the system.}---or if it truly require physics beyond GR.
We argue that this is potentially the most powerful tool for distinguishing waveform systematics from new physics~\citep[e.g., see][]{gupta_possible_deviations_24}.

The remainder of this paper is organised as follows. In Section~\ref{section:mathematical_foundation} we introduce our signal model. In Section~\ref{section:inverse_problem}, we outline how to apply this model to data, and show why a standard solution is computationally intractable. In Section \ref{section:formalism}, we summarise our method for solving the problem outlined in Section~\ref{section:mathematical_foundation}, using a pragmatic, two-step approach, combining the single-event Gaussian-process formalism of \citet{passenger_gaussian_2025} with a coherent combination approach for multiple events.
In Section~\ref{section:injection_study}, we validate our framework with analysis of synthetic gravitational-wave events. In Section \ref{section:waveform_systematics}, we show how our method can be used to distinguish between waveform systematics and a deviation from GR. In Section~\ref{section:GWTC-5}, we analyse 28 high SNR binary black hole events from GWTC-5 \citep[][]{collaboration_gwtc-50_2026}. We find no evidence for deviations from GR. In Section~\ref{section:discussion}, we discuss the impact of assumptions and approximations on our results, and outline directions for future work.
 
\section{Signal model}\label{section:mathematical_foundation}\label{section:mapping}

\subsection{The deviated-strain map}
Our signal model is described by a deviated-strain \textit{map}: 
\begin{align}\label{eq:map}
    s_{+,\times}(\tau , M_s) ,
\end{align}
which depends on the source-frame total mass $M_s$ and on ``natural time'' $\tau$:
\begin{align}
    {\tau} = \frac{c^3}{G M_s}t_s .
\end{align}
Here, $c$ is the speed of light, $G$ is the gravitational constant, and $t_s$ is the source-frame time relative to the time of coalescence.
Equivalently, we can write $\tau$ in terms of the redshifted detector-frame mass $M_\text{det}=(1+z) M_s$ and detector-frame time $t_\text{det} = (1+z) t_s$:
\begin{align}\label{eq:natural_time}
    {\tau} = \frac{c^3}{G M_\text{det}} t_\text{det} .
\end{align}
The map is the difference between the true strain $h_\text{true}$ and the one predicted from general relativity $h_\text{GR}$:
\begin{align}
    s_{+,\times}(\tau, M_s) \equiv 
    h_\text{true}^{+,\times}(\tau , M_s) - 
    h_\text{GR}^{+,\times}(\tau , M_s) .
\end{align}

\begin{figure*}
    \centering
    \includegraphics[width=18cm]{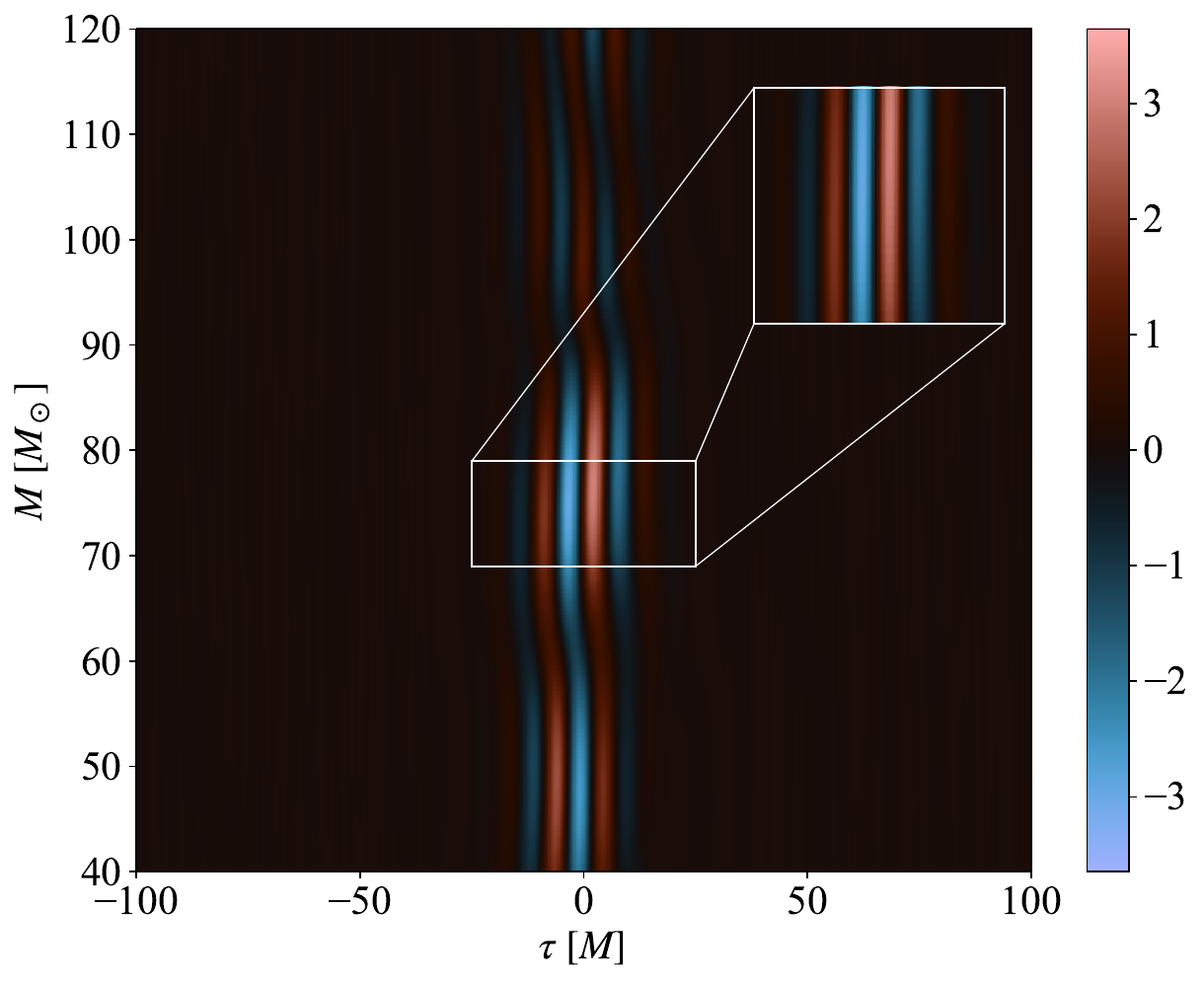}
    \caption{
    A visual representation of the deviated strain $s_{+,\times}(\tau, M)$.
    We show only the $+$ component, but there is another similar map for the $\times$ component.
    The horizontal axis is natural time in units of total mass relative to the coalescence.
    The vertical axis is the total mass.
    The color bar is the $+$ component of the deviated strain.
    In this example, the deviated strain is approximately coherent over scales of $10 M_\odot$, which is the size of the white box.
    Beyond this scale, the deviated strain is not coherent in mass.}
    \label{fig:2d_map_example}
\end{figure*}

We refer to Eq.~\ref{eq:map} as a ``map'' because one can represent $s$ as a two-dimensional plot. 
Figure \ref{fig:2d_map_example} shows an example visualization of one polarization component of the deviated strain: $s_+(\tau, M)$. 
In this example, the deviated strain is most pronounced at $\tau = 0$ and $M \approx 75 M_\odot$.
This map is characterized by $\gtrsim 8\times10^5$ parameters: 2048 bins in $\tau$, 200 bins for $M$, and two polarization states $i$.
We denote the set of these parameters as $\mathfrak{s}$ so that
\begin{align}
    \mathfrak{s}_{ijk} = s_i(\tau_j, M_{s,k}) .
\end{align}

In principle, the map can depend on other parameters besides $M_s$ such as mass ratio or black-hole spins.
For the sake of simplicity, however, we assume that the map depends only on $M_s$, and leave the inclusion of additional parameters for future work.

The map is defined for a fiducial observer who witnesses mergers at a fixed inclination $\iota=0$, polarisation angle $\psi=0$, coalescence phase $\phi_c=0$ and distance $d_L=\unit[1]{Mpc}$.
In subsection \ref{subsection:detector_frame}, we describe how to use the map to calculate the deviated strain for an arbitrary observer.

For simplicity, we introduce some shorthand that will help us to keep the notation relatively compact.
In the remainder of the manuscript, we drop the subscript $s$ and refer to source-frame mass as simply $M$.
Meanwhile, we drop the ``det'' from detector-frame time, which is denoted simply as $t$.

\subsection{Statement of the problem}
The goal of this paper is to estimate the parameters $\mathfrak{s}$ that characterize the deviated strain map.
These parameters uniquely determine the deviations from GR in each event.
Since the map is common to every event, data from multiple events can be combined to infer the map values.

\subsection{The detector frame}\label{subsection:detector_frame}
The map determines the beyond-GR signal for an observer measuring a binary with total mass $M$ at coordinates $\iota, \psi, \phi_c, d_L$.
The first step to calculate the deviated strain for an arbitrary observer is to find the row of the map corresponding to $M$ and its corresponding time series $s_{+,\times}(\tau)$ in natural time.
The next step is to use Eq.~\ref{eq:natural_time} to convert the time series to detector-frame time: $s_{+,\times}(t)$.
This results in the metric perturbation for an observer at the fiducial coordinates (but using a clock located at $d_L$).

The next step is to convert the metric perturbation to $s'(t)$---the one observed by an observer at $\iota, d_L, \phi_c$ (we get to $\psi$ in a moment).
The transformation is:
\begin{align}\label{eq:transformation}
     \begin{pmatrix}
     s_+' \\ 
     s_\times' 
     \end{pmatrix}
    = \frac{e^{2i\phi_c}}{d_L}
    \begin{pmatrix}
        \frac{1}{2}(1+\cos^2\iota) & 0 \\
        0 & \cos\iota
    \end{pmatrix} 
         \begin{pmatrix}
     s_+ \\ 
     s_\times 
     \end{pmatrix} .
\end{align}
Now we can determine the differential strain in a gravitational-wave observatory using the usual antenna factors
\begin{align}\label{eq:antenna_factors}
    h_\text{dev} = F^+(\hat\Omega, \psi, t_c) s_+ 
    +
    F^\times(\hat\Omega, \psi, t_c) s_\times ,
\end{align}
which depend on sky location $\hat\Omega$, polarization angle $\psi$, and time of coalescence $t_c$ in the detector frame.

\section{The inverse problem}\label{section:inverse_problem}
In this Section, we outline---in theory---how to solve the problem laid out in Section \ref{section:mapping}. 
We discuss why a standard solution is computationally challenging.
This motivates a more pragmatic approach, which is less computationally daunting if also less elegant. 
Some readers may wish to skip directly to Section \ref{section:formalism}, which describes the pragmatic approach we ultimately settle on for this paper.

We work with two-detector, frequency-domain quantities unless otherwise specified. 
To keep the equations simple, the frequency-dependence and observatory index are both implicit.
For a single event $i$, the data $d_i$ can be modeled as
\begin{align}\label{eq:data}
    d_i = h_\text{GR}(\theta_i) + h_\text{dev}(\theta_i, \mathfrak{s}) + n_i ,
\end{align}
where $h_\text{GR}(\theta_i)$ is the deviated strain due to GR, which depends on binary parameters $\theta_i$.
Meanwhile, $h_\text{dev}(\theta_i, \mathfrak{s})$ is the differential strain from the deviated strain, which depends on the total source-frame mass $M_i$.
Due to the antenna factors in Eq.~\ref{eq:antenna_factors}, $h_\text{dev}(\theta_i, \mathfrak{s})$ also depends on a few parameters in $\theta$: sky location $\hat\Omega$, polarization angle $\psi$, and time of coalescence $t_c$.
It also depends on the deviated-strain map parameters $\mathfrak{s}$.
Finally, $n_i$ is the detector noise.

The frequency-domain likelihood is then
\begin{widetext}\label{eq:likelihood}
    \begin{align}
    \mathcal{L}(d_i|\mathfrak{s},\theta_i) \propto \frac{1}{\det{N_i}}\exp\left[-\big(d_i - h_\text{GR}(\theta_i) - h_\text{dev}(\theta,\mathfrak{s})\big)^\dagger N_i^{-1} \big(d_i - h_\text{GR}(\theta_i) - h_\text{dev}(\theta,\mathfrak{s})\big) 
    \right].
\end{align}
\end{widetext}
Here, $N_i$ is the covariance matrix of the noise,
\begin{align}\label{eq:N}
    N_i \equiv & \langle n_i n_i^\dagger \rangle \nonumber \\
    = & \frac{1}{2\Delta f} \operatorname{diag}\left[P_i\right],
\end{align} 
where $\Delta f$ is the frequency-bin width and $P_i$ is the single-sided noise power spectral density (PSD).
This choice of convention for $N_i$ absorbs the usual factor $1/2$ in the Whittle likelihood, which allows for direct application of standard Gaussian-process identities in later sections.

The likelihood of observing the data $d$ for $Q$ events with binary parameters $\vec\theta$ given deviated-strain map parameters $\mathfrak{s}$ is:
\begin{align}
    \mathcal{L}(d|\vec\theta, \mathfrak{s}) = \prod_i^Q \mathcal{L}(d_i|\theta_i, \mathfrak{s}) .
\end{align}

The next step is to assume a prior for $\mathfrak{s}$.
Our prior belief is that deviations from GR should be similar for binaries with similar total mass.
We therefore adopt a prior that ensures that $s(\tau, M)$ varies slowly with $M$.
To that end, we assume a zero-mean Gaussian-process, which we denote:
\begin{align}\label{eq:full_kernel_prior}
    \pi(\mathfrak{s}\mid \Lambda) .
\end{align}
This is a multivariate normal distribution, with a covariance matrix $K(\Lambda)$ that depends on hyper-parameters $\Lambda$. 

We discuss the detailed functional form of $K$ in the next section.
For our present purposes, it is enough to say that $K$ is designed so that the deviated strain:
\begin{itemize}
\item is localised near the merger $(\tau=0)$;
\item oscillates at a characteristic frequency (controlled by $\Lambda$); and
\item is approximately unchanged in the $M$ direction over length scales of $\Delta M$, which is one of the hyper-parameters.
\end{itemize}

Putting everything together, one can write down a posterior distribution for the deviated-strain map parameters:
\begin{align}
        p(\mathfrak{s} | d) \propto 
        \int d\Lambda\, 
        \pi(\mathfrak{s} \mid \Lambda) \, \pi(\Lambda)
        \prod_i^Q 
        \int d\theta_i\, 
        {\cal L}(d_i | \theta_i, \mathfrak{s}) \, 
        \pi(\theta_i).
\end{align}

If we were able to make draws of $\mathfrak{s}$ from this posterior distribution, we could reconstruct the deviated-strain map.
However, this is quite challenging in practice.
To see why, we first make a couple of assumptions designed to make the problem \textit{easier}.
(If we can show this easier version of the problem is computationally difficult, we can be assured that the harder problem is even more challenging.)
First, we work in the high-SNR limit so that there are no integrals over $\theta_i$.
Second, we assume that the hyper-parameters $\Lambda$ are perfectly known so that the there is no integral over $\Lambda$ (in other words, the prior on $\Lambda$ is a delta function at the true value).
With these assumptions, the posterior becomes a multivariate normal distribution in $\mathfrak{s}$.
Since $\pi(\mathfrak{s})$ is a Gaussian and each ${\cal L}(d_i|\mathfrak{s})$ is a Gaussian in $\mathfrak{s}$, their product is also a Gaussian.

In theory, the above simplifications allow us to directly draw samples of $\mathfrak{s}$.
The covariance matrix for the posterior distribution is simply 
\begin{align}
    C = (K^{-1} + \Xi^{-1})^{-1} ,
\end{align}
where $K$ is the covariance matrix in the prior and $\Xi$ is the one obtained from the product of likelihoods.
This is where things break.
Since there are $8\times10^5$ pixels in $\mathfrak{s}$, there are ${\cal O}(10^{10})$ elements in $K$ and $\Xi$.
These matrices cannot be held in computer memory.\footnote{When we created Fig.~\ref{fig:2d_map_example}, we took advantage of the fact that the generative model is a \textit{separable} Gaussian process so we could split the calculation into two steps. However, this separability is spoiled when $\mathfrak{s}$ is conditioned on the data.}
And they are far too large to be inverted.
Thus, a straightforward solution with standard methods is computationally difficult.

To make the problem tractable, we approximate the full covariance treatment of $K$ using a two-step process:
\begin{enumerate}
    \item We estimate the kernel hyper-parameters $\Lambda$ while assuming that the length scale in the $M$ direction $\Delta M\rightarrow0$. This simplifies the problem immensely --- each event is now an independent draw of the same process.
    \item We reintroduce correlations due to $\Delta M$ in post-processing, where we assume $\Delta M \neq 0$ and use it to correlate nearby events, such that events with identical $M$ have identical deviated strain, and the deviated strain decorrelates between events over a scale $\Delta M$.
\end{enumerate}
These two steps are not internally consistent, as they implicitly ignore any correlation between $\Delta M$ and the other hyperparameters.
However, they allow us to develop a sub-optimal estimator, which is adequate to get things off the ground.
It provides a qualitatively correct reconstruction of the deviated-strain map, albeit, with less statistical power than we would have from an optimal solution.
In the next section we develop this compromise solution.

\section{Formalism}\label{section:formalism}
Here, we outline the (sub-optimal) method for solving the problem described by Section \ref{section:mapping}. In Subsection \ref{subsection:GP_formalism}, we build a framework for estimating the shared parameters for the deviated strain across gravitational-wave observations, while assuming independence. In Subsection \ref{subsection:total_mass_correlation}, we show how to reintroduce coherence between events, by assuming that events with difference in total mass $\lesssim\Delta M$ should have similar deviated strains.

\subsection{Single-event analysis}\label{subsection:GP_formalism}
We begin by outlining the single-event Gaussian process framework of \citet[][]{passenger_gaussian_2025}. We work with data from two gravitational wave-detectors, $\mu$ and $\nu$. As noted in Section \ref{section:inverse_problem}, the differential gravitational-wave strain is 
\begin{align}\label{eq:data}
    d_i = h_\text{GR}(\theta_i) + h_\text{dev}(\theta_i, \mathfrak{s}) + n_i ,
\end{align}
where $d_i$ is the data, $h_\text{GR}(\theta_i)$ is a waveform approximant, $h_\text{dev}(\theta_i,\mathfrak{s})$ is the deviated strain and $n_i$ is the detector noise. 

We make a number of assumptions:
\begin{enumerate}
    \item We make the approximation that $h_\text{GR}(\theta_i)$ is adequately specified, which is to say we ignore waveform error.
    \item We assume that the SNR is sufficiently high that we know $\theta_i$, thereby allowing us to perfectly subtract the GR waveform and work with residual data $r_i=d_i-h_\text{GR}(\theta_i)$.
    \item We assume that $|h_\text{dev}(\theta_i, \mathfrak{s})|$ is small enough compared to $|h_\text{GR}(\theta_i)|$, such that it does not significantly affect the usual single-event estimation of the binary parameters $\theta_i$.
\end{enumerate}
We do not expect these assumptions to be entirely valid---especially $(2)$ and $(3)$---but we make them anyway in order to provide a starting point from which we can sketch out the rest of the formalism.
We return to these assumptions later, testing how they work with simulated data and sketching out how we might construct a more robust framework without these assumptions.

We consider slices of fixed $M$ values in $\mathfrak{s}$.
Given a fixed $M$, we assume that $\mathfrak{s}$ can be modeled as a zero-mean Gaussian process.
We assume that the deviated strain is localised in time near the merger and that it oscillates with some characteristic frequency. We do not assume it to be symmetric about the merger time. 
We assume that each polarisation state is described by a $\tau$-domain kernel matrix $K^\tau$ with hyper-parameters $\Lambda$.
We assume the $+$ and $\times$ polarisations behave the same way on average, so that
\begin{align}
    \langle \mathfrak{s}^{+\dagger}_{a} \mathfrak{s}^+_{b}\rangle = \langle \mathfrak{s}^{\times\dagger}_{a}
    \mathfrak{s}^\times_{b} \rangle \equiv K^\tau_{ab}(\Lambda) .
\end{align}
In other words, we assume the $+$ and $\times$ are independent realisations of the same underlying process.

We choose the following form for $K^\tau$:
\begin{widetext}
    \begin{align}\label{eq:Ktau}
        K^\tau_{ab}(\Lambda) \equiv & \
        k_0 \exp{\left(-f_0^2\frac{\tau_a^2+\tau_b^2}{2 w^2}\right)}
        \exp{\left(-f_0^2\frac{(\tau_a - \tau_b)^2}{2l^2}\right)}
        \cos\left(2\pi f_0 |\tau_a - \tau_b|\right) .
    \end{align}
\end{widetext}
There are four Gaussian-process hyper-parameters in $\Lambda$:
\begin{itemize}
    \item A scale parameter $k_0$, in units of strain, which controls the amplitude of the deviated strain at the merger time.
    \item The characteristic frequency of the deviated strain $f_0$ in units of inverse natural time.
    \item The characteristic duration $w$ (unitless).
    \item The unitless coherence time $l$. 
\end{itemize}

The hyper-parameters $\Lambda$ describe the correlations between natural time points $(\tau_a, \tau_b)$ for the fiducial observer at $d_L^\text{ref}=\unit[1]{Mpc}$. While the Gaussian process is the same for all events when expressed in units of natural time $\tau$ (with a single $\Lambda$ value), the $\Lambda$ value \textit{maps differently to detector-frame time for each event}, depending on the detector-frame total mass and luminosity distance of the event (see Eq.~\ref{eq:natural_time}). 
In particular,\footnote{As $w$ and $l$ are dimensionless parameters, they are not affected by the transformation from natural time to dimensional time.}
\begin{align}\label{eq:lambda_k}
    k_0 \rightarrow & \frac{k_0}{d_L^2} \\
    f_0 \rightarrow &  f_0 \frac{c^3}{GM_\text{det}} .
\end{align}
As a consequence, the characteristic frequency in Hz is lower for events with higher $M_\text{det}$---even though they share the same hyper-parameters, which are defined in terms of natural time for the fiducial observer.
The revised kernel (in detector frame units) is denoted $K^t_i$ instead of $K^\tau$.
For the remainder of the subsection, we work with quantities measured in detector-frame units.

We transform $K^t_i$ from the time domain to the one-sided frequency domain, using a one-sided DFT matrix:
\begin{align}
    K^f_i = U K^t_i U^{\dagger} .
\end{align}
Here,
\begin{align}\label{equation:dft_matrix}
    U_{ab} \equiv & \Delta t\, \exp(-2 \pi \mathrm{i}\, f_a t_b) ,
\end{align}
where $\Delta t$ is the spacing between time points.
Henceforth, we work with the frequency-domain kernel matrix $K^f_i$, and drop the superscript $f$.

We can now write an expression for $S_i$---a matrix describing the covariance between the differential strain in detectors $\mu$ and $\nu$:
\begin{widetext}
    \begin{align}\label{eq:Sij}
    S_i \equiv &
    \left(
    \begin{matrix}
    \langle{{h_{\text{dev},i}^\mu}^\dagger h_{\text{dev},i}^\mu}\rangle & \langle{{h_{\text{dev},i}^\mu}^\dagger h_{\text{dev},i}^\nu}\rangle \\
    \langle{{h_{\text{dev},i}^\nu}^\dagger h_{\text{dev},i}^\mu}\rangle & \langle{{h_{\text{dev},i}^\nu}^\dagger h_{\text{dev},i}^\nu}\rangle
    \end{matrix}
    \right)
    \nonumber \\
    = &
    \left(
    \begin{matrix}
    \left(F_{\mu,i}^{+2} A_{+,i}^2 + F_{\mu,i}^{\times 2} A_{\times,i}^2\right) K_i &
    (F_{\mu,i}^+ F_{\nu,i}^+ A_{+,i}^2 + F_{\mu,i}^\times F_{\nu,i}^\times A_{\times,i}^2) K_i D_i^\dagger \\
    (F_{\mu,i}^+ F_{\nu,i}^+ A_{+,i}^2 + F_{\mu,i}^\times F_{\nu,i}^\times A_{\times,i}^2) D_iK_i &
    \left(F_{\nu,i}^{+2} A_{+,i}^2 + F_{\nu,i}^{\times 2} A_{\times,i}^2\right) D_iK_iD_i^\dagger
    \end{matrix}
    \right) .
\end{align}
\end{widetext}
Here, $D_i$ is a frequency-dependent phase factor. It accounts for the time delay between detectors $\delta{t}_{\mu\nu}$, which depends on the sky location $\hat\Omega$,
\begin{align}\label{eq:D_matrix}
    D_i \equiv \operatorname{diag}\left[e^{-2\pi \text{i} {f} \delta{t}_{{\mu\nu,i}}}\right] .
\end{align}

The likelihood for residual data $r_i$ in a single event is
\begin{align}\label{eq:single_event_likelihood}
    \mathcal{L}(r_i \mid \Lambda_i)
    &= \frac{1}{\pi^n \det{{(S_i + N_i)}}}\exp{\left(-r^\dagger_i {(S_i+N_i)}^{-1}r_i\right)}.
\end{align}

If we were building the generative model outlined in Section \ref{section:inverse_problem}, now is when we would introduce our prior belief that the deviated strain is the same in events with the same total mass $M$, and decorrelates between events over some distance scale. This could be achieved by introducing a new kernel function that correlates nearby total-mass points. However, as we showed above, such an approach is computationally infeasible. Therefore, for now, we ignore correlations between events, and reintroduce them in Subsection \ref{subsection:total_mass_correlation}.

Ignoring correlations between events is equivalent to assuming that $\Delta M \rightarrow 0$, so that each event's deviated strain is an independent draw from the same process. The full kernel matrix $K_\text{full}$ from Section \ref{section:inverse_problem}, which includes a non-zero $\Delta M$, can be written as a block matrix with non-negligible off-diagonal terms. Assuming $\Delta M \rightarrow 0$ allows the inverse of this matrix to be treated as block-diagonal, decomposing into independent, per-event inversions. This means we can get a point estimate for $\Lambda$, which we later use to combine events when we reintroduce $\Delta M$. The joint likelihood is the product over events,
\begin{align}\label{eq:joint_likelihood}
    \mathcal{L}(r \mid \Lambda) = \prod_i^Q \mathcal{L}(r_i \mid \Lambda_i) .
\end{align}  
The posterior for $\Lambda$ given $r$ is 
\begin{align}\label{eq:joint_posterior}
    p(\Lambda \mid r) \propto\prod_i^Q \left[\mathcal{L}(r_i \mid \Lambda_i)\right]  \pi(\Lambda),
\end{align}  
where $\pi(\Lambda)$ is the prior for $\Lambda$.
We denote the maximum-likelihood estimate of the hyper-parameters as $\widehat{\Lambda}$.

\subsection{Total mass correlations}\label{subsection:total_mass_correlation}
Now that we have an estimate for the Gaussian process hyper-parameters $\widehat\Lambda$, we reintroduce correlations due to $\Delta M$ using the deviated strain estimates for each event to construct the map of $s_{+,\times}(\tau , M)$. However, this two-step scheme requires two additional considerations:
\begin{enumerate}
    \item The deviated strain must be evaluated on common natural-time grid points, since events of different detector-frame total mass are stretched by different amounts.
    \item Each event's deviated strain must be transformed to a fiducial frame, since events are observed at different sky locations, distances and orientations.
\end{enumerate}

To address the first consideration, we make use of standard Gaussian-process identities \citep[see, e.g.,][]{rasmussen_gaussian_2008}. First, we relabel the time axis for each event from dimensional time to natural time (see Eq.~\ref{eq:natural_time}),
\begin{align}
    \tau_i = \frac{c^3}{G M_{\text{det},i}}t_i.
\end{align}
Then, the predicted posterior mean for the deviated strain can be calculated as 
\begin{align}\label{eq:predicted_mean}
    s'_{\mu,\nu,i}(\tau , M) =B_i  \left(S_i + N_i\right)^{-1} r_i ,
\end{align}
where $B_i$ is a rectangular matrix which interpolates from the implicit $f_{\tau,i}$ grid, with spacing\footnote{The $\unit[0.5]{Hz}$ comes from the fact that we are looking at $\unit[2]{s}$ segments (see Section \ref{section:injection_study}).}

\begin{align}
    \Delta f_i = \frac{G M_{\text{det},i}}{c^3} \unit[0.5]{Hz} ,
\end{align}
to the $\tau$ grid we are using for the map, with unitless spacing of 1 (for additional details, see Appendix \ref{appendix:choosing_tau}).
The matrix $B_i$ is calculated by forming each $K^\tau$ in Eq.~\ref{eq:Sij} between $\tau_i$ and $\tau$ using Eq.~\ref{eq:Ktau}, restricted to the natural-frequency bins that fall within event $i$'s own observing band (see Appendix \ref{appendix:choosing_tau})  followed by the appropriate transformations. 

The covariance of the deviated strain on the common $\tau$ grid is 
\begin{align}\label{eq:predicted_covariance}
    \Sigma_{\mu,\nu,i}' = S_{{**},i} - B_i \left(S_i + N_i\right)^{-1} B_i^\dagger ,
\end{align}
where $S_{{**},i}$ is constructed analagously to $B_i$, except between pairs of $\tau$ points.

To transform these quantities to a common observing frame, we perform the inverse of the transformation in Eq.~\ref{eq:transformation}, so that the mean predicted deviated strain is
\begin{align}
    \bar{s}_i = & T_i^{-1} s_i' ,
\end{align}
where 
\begin{align}
    T_i = \frac{e^{2i\phi_{c,i}}}{d_{L,i}}
    \begin{pmatrix}
        \frac{1}{2}(1+\cos^2\iota_i) & 0 \\
        0 & \cos\iota_i
    \end{pmatrix} 
\end{align}
Transforming the covariance matrix to the fiducial observer, we obtain
\begin{align}
    \Sigma_i = T_i^{-1} \Sigma_i' (T_i^{-1})^\dagger .
\end{align}

Now, to reintroduce correlations between events, we use the information from all events to predict the deviated strain at total-mass index $\alpha$ using a weighted average,
\begin{align}\label{eq:stacking_for_map}
    \hat{s}_{i} = \left(\sum_{j}^Q w_{ij}(\Delta M)\, \Sigma_j^{-1}\right)^{-1}\left(\sum_{j}^Q w_{ij}(\Delta M) \, \Sigma_j^{-1} \, \overline{s}_j\right) .
\end{align}
Here, $\hat{s}_{i}$ is an estimator for the deviated strain at total mass index $i$, and $\Sigma_j^{-1}$ is its Fisher information matrix, which encodes the uncertainty in its estimate. Meanwhile, $w_{ij}(\Delta M)$ are weights which depend on $\Delta M$,

\begin{align}\label{eq:weight}                           
  w_{ij}(\Delta M)= \exp\bigg(-\frac{1}{2}\frac{(M_i-M_j)^2}{\Delta M^2} \bigg),
\end{align}  
which determines the degree of correlation between total mass index $i$ and event $j$. In this scheme, events with the same total mass \textit{have the same deviated strain}, and their deviated strain decorrelates over some scale $\Delta M$.

Calculation of the quantity $\Sigma_i^{-1}$ can become ill-conditioned because, by construction, $\Sigma_i$ assigns negligible variance to deviations far from its characteristic frequency and duration. 
This can lead to spuriously large values when the matrix is inverted. To mitigate this, we perform a singular-value decomposition,
\begin{align}
    \Sigma_i = V_i \lambda_i V_i^\dagger ,
\end{align}
where $V_i$ is a matrix whose columns correspond to the eigenvectors of $\Sigma_i$, and $\lambda_i$ are the eigenvalues of $\Sigma_i$. Then, we regularise by flooring each $\lambda_i$ at a fraction $\gamma$ of the largest eigenvalue. We choose $\gamma=10^{-6}$, which seems to strike a balance between increasing uncertainty in the zero-variance, prior-dominated regions of the signal, while not vastly increasing uncertainty where the deviated strain is $\neq 0$. Formally, we floor eigenvalues as 

\begin{align}\label{eq:safe_precision}
    \tilde\lambda_{i,k} = \max\left(\lambda_{i,k},\ \gamma\lambda_{i,\text{max}}\right) ,
\end{align}
then the inverse of $\Sigma$ becomes
\begin{align}
    \Sigma_i^{-1} \approx V_i\tilde\lambda_i^{-1}V_i^\dagger .
\end{align}

\subsection{Event selection criteria}\label{subsection:event_selection_criteria}

Our framework for combining events assumes each event has a well-measured total mass. Events with large total-mass uncertainty $\sigma_M$ blur the correlation structure of the deviated strain. If $\sigma_M \gtrsim \Delta M$, the event cannot be meaningfully localised in total mass space. 
If one used the generative model described in Section \ref{section:inverse_problem}, in principle, marginalisation over the total mass uncertainty for each event is possible. However, in our framework, it is not straightforward to marginalise over total mass uncertainty in a rigorous way.

Instead, we impose a selection criterion via network SNR to retain only the most informative events. We employ a cut of network SNR $\gtrsim 15$. 
The median $90\%$ credible interval source-frame total mass posterior width for these events is $\approx 11 M_\odot$.
We also choose to only analyse events with dominant ringdown frequencies inside the range $[\unit[50]{Hz},\unit[511]{Hz}]$ as calculated using the \textsc{qnm} package \citep[][]{stein_qnm_2019}. Finally, we do not include events with source-frame total mass $\gtrsim 120\,M_\odot$, as the sparsity of these events means their deviated strain is poorly informed by neighbouring events, and therefore do not contribute meaningfully when events are combined. This leaves $28$ events (see Appendix \ref{appendix:events_analysed}).

\section{Tests with simulated data}\label{section:injection_study}
We test the analysis to show that we recover a deviated strain signal when we know one is present. 
In doing so we demonstrate that the approximations employed in Section~\ref{section:formalism} do not prevent us from obtaining a qualitatively accurate reconstruction of the deviated strain.

We use a synthetic dataset consisting of 200 gravitational-wave events, with the Hanford and Livingston gravitational-wave detectors.
Each data segment is $\unit[2]{s}$ long with a frequency range of [$\unit[50]{Hz},\unit[511]{Hz}$] and a sampling frequency of $\unit[1024]{Hz}$.
The lower bound is chosen due to the possible presence of non-Gaussian noise below $\unit[50]{Hz}$ \citep[see, e.g.,][]{davis_ligo_2021, macas_sensitive_2023, cheung_does_2024, soni_ligo_2025}.
For each event, we have three data components: detector noise $n$, a gravitational-wave signal described by GR $h_\text{GR}$ and deviated strain $s$.

For each event, we fetch $\unit[130]{s}$ of data at a random time from the second part of the third observing run \citep[O3b,][]{abbott_open_2023}, with GPS time in the range $[1256655618, 1269363618]$. We check each data segment does not overlap with any glitches identified in the \textsc{GravitySpy} database~\citep[][]{zevinGravitySpyIntegrating2017} with more than $50\%$ confidence. To model $n$, we estimate the PSD using Welch's method on the first $\unit[128]{s}$ of data, with a segment length of $\unit[2]{s}$ and a Tukey window with roll off $\alpha=0.4$. We use the median value for each frequency bin. The remaining $\unit[2]{s}$ of data is used as the noise $n$ for the event.

For $h_\text{GR}$, we assign each event a source-frame total mass equally spaced in the range $[40 M_\odot, 120M_\odot]$, with $m_1=m_2$, and zero spin magnitudes. The remaining parameters are drawn from standard priors, except for luminosity distance, which is chosen such that events have network SNR of $20$. We assume that GR is well described by the \textsc{IMRPhenomPv2} waveform \citep[][]{schmidtModelsGravitationalWaveforms2012}.

For $s$, we draw deviated strain at the total mass of each event from a two-dimensional Gaussian process,\footnote{As our framework for coherently combining events only defines an estimator, rather than a generative model, we use a structurally similar two-dimensional Gaussian process as a stand-in.} in which one axis is natural time $\tau$ (in units of $M$), and the other is the source-frame total mass $M$. 
For additional details, see Appendix \ref{appendix:data_generation}. We use injection hyper-parameter values $k_0 = 2.8\times10^{-39}, \ w=\unit[3.0]{cycles}, \ l=\unit[2.0]{cycles}$ and $f_0= 0.075 \ M^{-1}$. We use a length scale in the mass direction $l_M$ of $10 M_\odot$. 
The mean network SNR of the deviated strain in each event is $2.0$.
We choose this value so that the deviations from GR are not obvious in any one event, but at the same time, the signal is loud enough that we can see it by stacking 200 events.

We estimate the posterior for the gravitational-wave binary parameters using the nested sampler \textsc{dynesty} \citep[][]{speagleDynestyDynamicNested2020}, implemented in \textsc{bilby}~\citep{ashtonBilbyUserfriendlyBayesian2019,bilby_gwtc1}. We use the \textsc{IMRPhenomPv2} waveform. We assume zero spins, and use a detector-frame chirp-mass prior covering the range of injected values with a $10\%$ buffer above and below, and a mass-ratio prior that is uniform on the interval $(0.05,\ 1)$. We use a uniform-in-comoving-volume distance prior, centered at the true luminosity distance for each event, spanning a factor of $100$ around it. We use standard priors on the remaining parameters. We use a small number of live points, $\textsc{nlive}=50$, as we are only interested in the maximum-likelihood binary parameter values. 

We subtract the maximum-likelihood waveform for each event to obtain residual data. We calculate the deviated-strain maximum a posteriori (MAP) Gaussian-process hyper-parameters\footnote{We do not perform Gaussian-process maximum-likelihood estimation, as it is ill-posed \citep[see, e.g.,][]{karvonen_maximum_2023}. This means regularisation via an informative prior distribution is necessary, even though such priors are difficult to motivate physically.} $\Lambda$ using Eq.~\ref{eq:joint_posterior}. We choose priors on these hyper-parameters in the detector-frame total-mass scaled frame such that they encompass a range of physically plausible merger-centric deviations:
\begin{itemize}
    \item \textbf{Amplitude:} $k_0$ is drawn from a log-uniform distribution with bounds of $[10^{-40},\ 10^{-37}]$.
    The upper bound of this range corresponds to the point at which deviated strain would likely have already been found in current analyses, while the lower bound is chosen to comfortably contain deviated strain with undetectably-low SNR. Given random draws from the prior on $f_0$ (since $w$ and $l$ are fixed), $k_0 = 10^{-37}$ produces deviated strain with a mean network SNR of $\approx{42}$ in the loudest event, GW250114, while $k_0 = 10^{-40}$ corresponds to a mean network SNR of $\approx{1}$.
    \item \textbf{Envelope and coherence:} We choose fixed values of $w=3.0$ and $l=2.0$. After testing several informative priors, we found that in the low-SNR signal limit, these parameters are poorly informed by the data, with $w$ drifting towards the upper bound of its allowed range, and $l$ drifting towards the lower bound. This lets the model overfit noise fluctuations, rather than signal structure. We fix $(w,l)$ = $(3.0,2.0)$ as these values are large enough to discourage the Gaussian process from overfitting noise, while enforcing our prior belief that the deviation is unlikely to be visible in band, nor remain phase coherent, for more than a few cycles. Choosing a smaller value of $l$ than $w$ allows the signal to be phase incoherent over its typical duration.
    \item \textbf{Frequency:} $f_0$ is drawn from a log-normal distribution, with a mean of $\ln\sqrt{f_\text{lo}f_\text{hi}}$ and a variance of $\left[\tfrac{\ln(f_\text{hi}/f_\text{lo})}{8}\right]^2$, truncated to 
    \begin{align}
        [f_\text{lo},\,f_\text{hi}] = [50\,\beta_\text{max},\ 511\,\beta_\text{min}] .
    \end{align}
    Here, $\beta_i \equiv GM_{\text{det},i}/c^3$ and $\beta_\text{max},\beta_\text{min}$ correspond to the highest and lowest detector-frame total-mass events respectively. This ensures no event's deviated-strain covariance matrix is defined outside the LIGO observing band. For our mass range, this gives $f_0 \in [0.046,\ 0.108]$.
\end{itemize}

We use the \textsc{L-BFGS-B} optimiser implemented in \textsc{scipy} \citep[][]{virtanen_scipy_2020} to calculate the MAP hyper-parameters for the deviated strain. We discuss calculation of analytic gradients for our model in Appendix \ref{appendix:gradients}. We calculate the posterior probability at 64 initialisations randomly drawn from the prior on each hyper-parameter, and optimise from the location of highest probability. 
The best-fit MAP hyper-parameters are $k_0=4.5\times 10^{-39}, w=\unit[3.0]{cycles}, l=\unit[2.0]{cycles}$ and $f_0= 0.074M^{-1}$. Using these, we construct the predicted mean deviated strain estimate for each event  $\bar{s}_{i}$ using Eq.~\ref{eq:predicted_mean}, as well as its predicted covariance matrix $\Sigma_i$ Eq.~\ref{eq:predicted_covariance}. 

\begin{figure*}
    \centering
    \includegraphics[width=18cm]{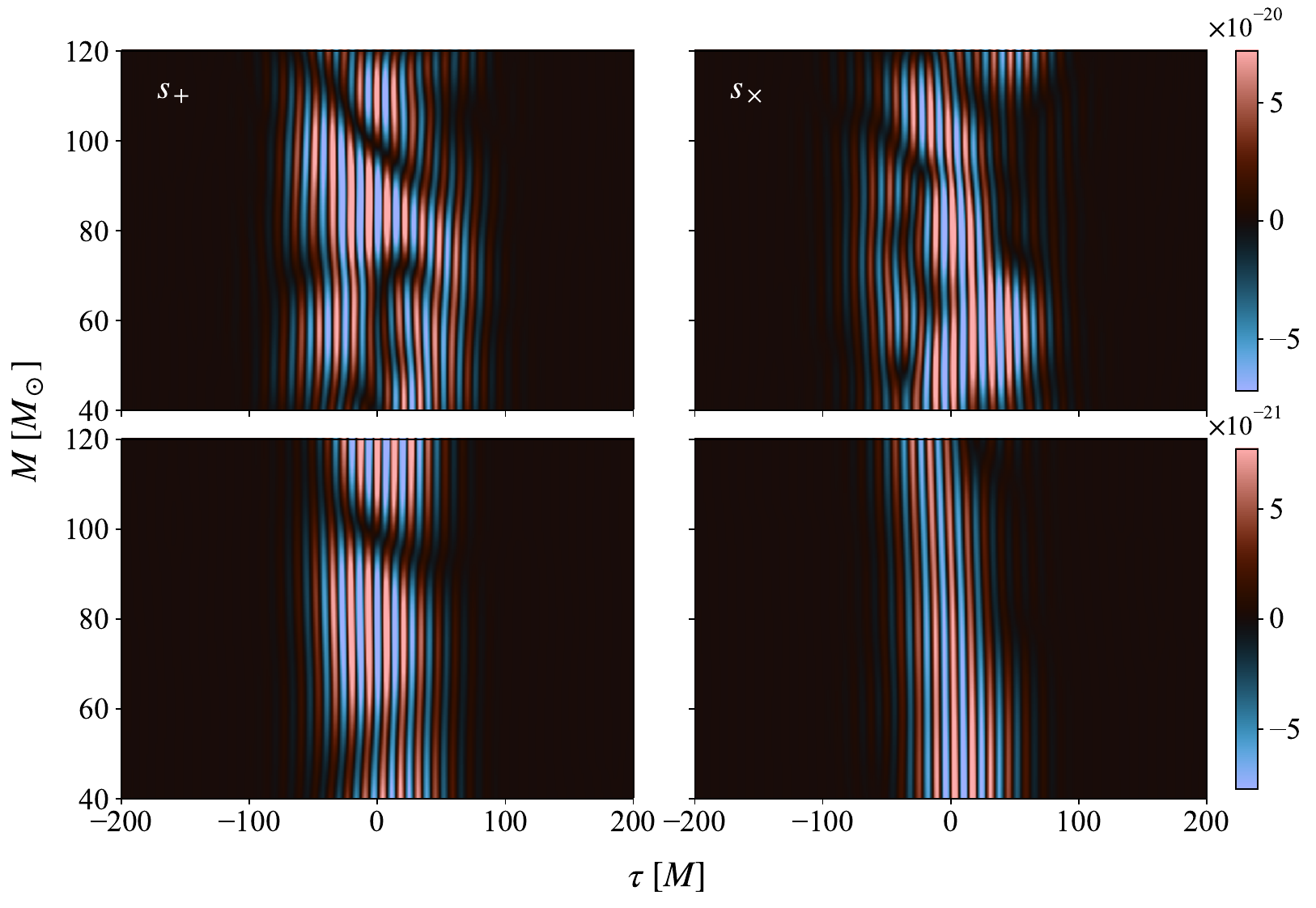}
    \caption{Heatmap for recovery of an injected deviated strain signal $\hat{s}$. The top left panel shows the injected coherent-frame deviated-strain $+$ polarisation, and the top right panel shows the $\times$ polarisation. The bottom two panels show their respective recovery $\hat{s}$. Qualitatively, there is broad agreement between the shapes of the injected deviated strain and the recovered $\hat{s}$.
    }
    \label{fig:2d_gp_injection}
\end{figure*}

We choose a $\Delta M$ value of $10 M_\odot$ and use Eq.~\ref{eq:stacking_for_map} to obtain $\hat{s}$. We show the resulting map of $\hat{s}$ in Figure \ref{fig:2d_gp_injection}, with the simulated signals in the top panels and the recovered signals in the bottom panels. We find broad agreement between the shape of the injected deviated strain map and the recovered map. The recovery of the $\times$ polarisation is worse than the $+$ polarisation, as we have defined it to be systematically weaker in the detector (see Eq.~\ref{eq:transformation}).
This gives us confidence that the approximations introduced in Section~\ref{section:formalism} still allow us to obtain qualitatively accurate reconstructions of the deviated strain.
The loss in signal power is due to the averaging of many noisy single-event estimates, which drives the overall signal estimate towards the prior mean of zero. 

In order to answer the question, ``is the structure in $\hat{s}$ statistically significant?" we define a detection statistic, $\lambda_\text{pop}$.
This is the likelihood ratio comparing the hypothesis that $\hat{s}>0$ to the hypothesis that $\hat{s}=0$. We derive an expression for $\widehat\lambda_\text{pop}$ in Appendix \ref{appendix:detection_statistic}. 
For this simulated data set, we obtain $\ln\widehat\lambda_\text{pop} = 9.49$, strongly supporting the presence of non-zero deviated strain. We discuss the significance of the value of $\widehat\lambda_\text{pop}$ in Appendix \ref{appendix:background_study}, where we calculate a $p$ value by comparing it to an empirical background distribution $\pi(\lambda_\text{pop}\mid \mathcal{H}_0)$. We find $p\approx0$, confirming recovery of a deviated strain signal.

\section{Identifying waveform-systematics}\label{section:waveform_systematics}
One advantage of our framework is its ability to identify spurious deviations from GR due to waveform systematics. As GR is scale invariant, waveform systematics appear as vertical bands in our deviated strain maps. In other words, these signals appear as signals with infinite length scale in the total mass direction, $\Delta M \rightarrow \infty$. If our framework finds evidence for a deviated strain signal favoring $\Delta M \rightarrow \infty$, waveform systematics are a likely explanation.

\begin{figure*}
    \centering
    \includegraphics[width=18cm]{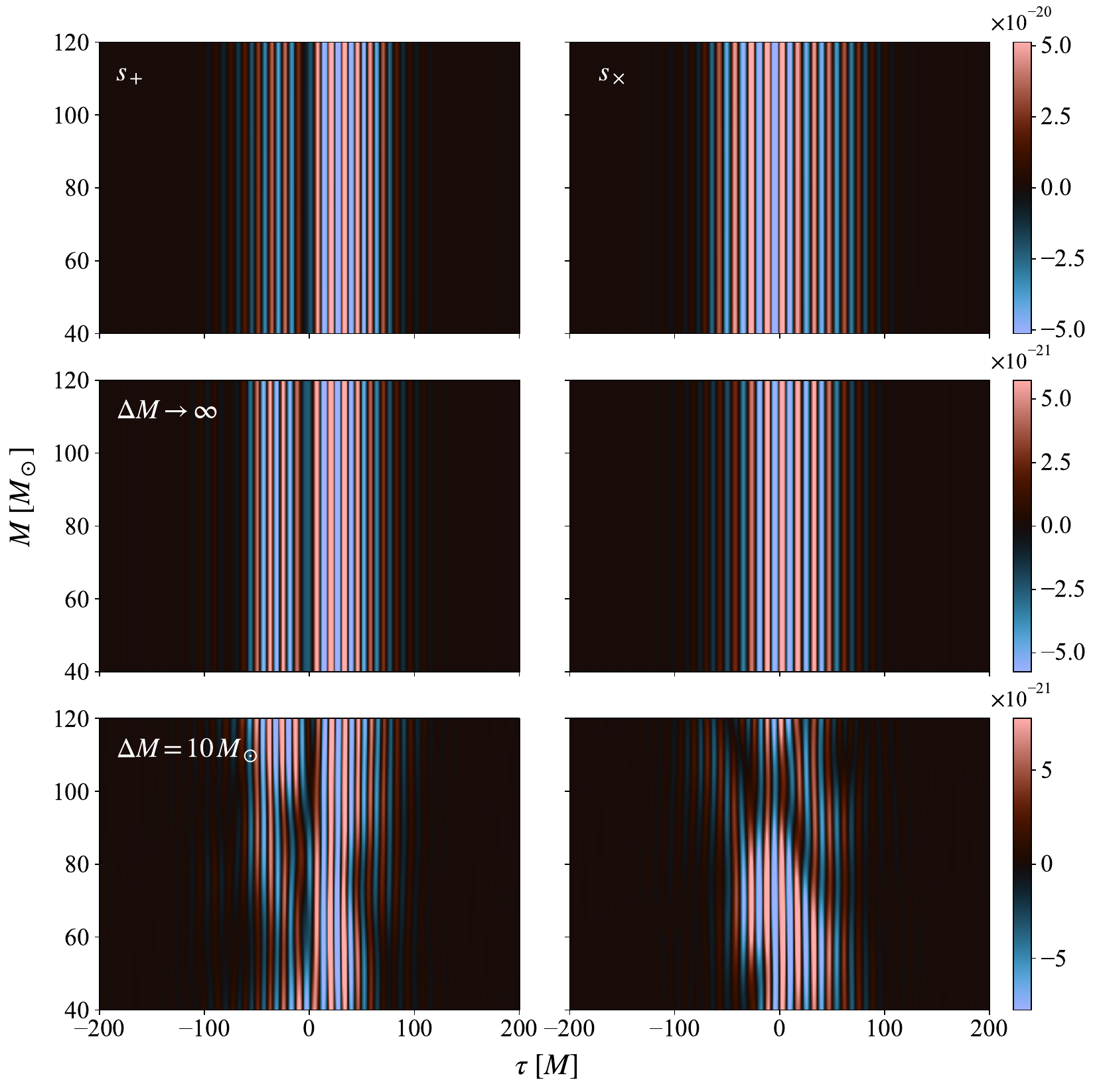}
    \caption{Heatmap for recovery of an injected strain signal consistent with waveform systematics. The top row shows the injected strain for the $+$ and $\times$ polarisations, the middle row shows the recovery using $\Delta M \rightarrow \infty$, and the bottom row shows the recovery using $\Delta M = 10 M_\odot$. 
    The data are better fit by the $\Delta M \rightarrow \infty$ model.
    }
    \label{fig:waveform_systematics_map}
\end{figure*}

To examine our framework's ability to identify waveform systematics, we perform the same injection study as in Section \ref{section:injection_study}, except we inject a deviated strain map that is mass-invariant (like we would expect from waveform systematics). We recover this injection using two analyses: one with $\Delta M = 10 M_\odot$, and one assuming no mass-dependence ($\Delta M \rightarrow \infty)$. The results for each analysis are shown in Fig.~\ref{fig:waveform_systematics_map}. 

We can calculate the significance for the $\Delta M\rightarrow \infty$ hypothesis by calculating the quantity
\begin{align}
    \mathcal{\lambda_\infty} = \frac{\mathcal{\lambda}_\text{pop}^{\Delta M\rightarrow \infty}}{\mathcal{\lambda}_\text{pop}^{\Delta M = 10}} ,
\end{align}
which is the likelihood ratio between each hypothesis, and each $\lambda_\text{pop}$ is calculated following the steps outlined in Appendix \ref{appendix:detection_statistic}. A value of $\mathcal{\lambda_\infty} >1$ indicates greater support for a mass-invariant signal, while $\mathcal{\lambda_\infty} <1$ supports the presence of a mass-dependent signal. 
For the example shown in Fig.~\ref{fig:waveform_systematics_map}, we find $\mathcal{\lambda_\infty}=1.6$, indicating support for the mass-independent model.
If we found such a result with real data, we would interpret this to mean that the deviated strain is caused by waveform systematics rather than a deviation from GR. In contrast, the mass-dependent injection from Section \ref{section:injection_study} has $\mathcal{\lambda_\infty}=0.01$, providing support for the new-physics hypothesis over the waveform systematic hypothesis.\footnote{In principle, one could calculate a $p$-value to determine the probability of obtaining a value of $\mathcal{\lambda_\infty}$. However, this would require a significant injection study, which we do not perform here.}

\section{Results for GWTC-5}\label{section:GWTC-5}
We analyse $28$ high SNR binary black hole events from GWTC-5 (see Section \ref{subsection:event_selection_criteria} for event selection criteria). Our analysis procedure is similar to Section \ref{section:injection_study}. We use data from the two gravitational-wave detectors Hanford and Livingston. For each event, we use the maximum likelihood parameters, which we obtain from public data releases \citep[][]{rich_abbott_open_2021, abbott_open_2023, collaboration_open_2025, collaboration_open_2026} of each catalog \citep[][]{the_ligo_scientific_collaboration_and_the_virgo_collaboration_gwtc-21_2024, ligo_scientific_collaboration_gwtc-3_2023, collaboration_gwtc-40_2025, collaboration_gwtc-50_2026}. For events from GWTC-1, GWTC-2 and GWTC-3, we use the \textsc{IMRPhenomXPHM} results, and for GWTC-4 and GWTC-5, we use the \textsc{IMRPhenomXPHM-SpinTaylor} results. Source-frame total masses are calculated assuming a flat $\Lambda$CDM cosmology with $H_0=\unit[67.9]{km/s/Mpc}$ and $\Omega_m=0.31$, consistent with the Planck 2015 results \citep[][]{ade_planck_2016}. 

We obtain $\unit[130]{s}$ of data from the Gravitational-Wave Open Science Centre \citep[GWOSC,][]{rich_abbott_open_2021, abbott_open_2023, collaboration_open_2025, collaboration_open_2026} with a frequency range of [$\unit[50]{Hz},\unit[511]{Hz}$] and a sampling frequency of $\unit[1024]{Hz}$, with the final $\unit[2]{s}$ of data centered at the maximum likelihood estimated merger time of each event. The first $\unit[128]{s}$ of data is used for the same noise PSD calculation method as in Section \ref{section:injection_study}, while the remaining $\unit[2]{s}$ is used as the data for the event. We choose not to use glitch-subtracted data frames, nor notch instrumental lines. Glitches in individual events are poorly estimated by neighbouring events and thus do not combine coherently. Poorly estimated instrumental lines appear at a common \textit{physical} frequency, and therefore do not stack coherently when we work with frequency in units of $M^{-1}$.

We subtract the waveform corresponding to the maximum-likelihood binary parameters for each event, and analyse the residual data using the procedure outlined in Section \ref{section:injection_study} to calculate the MAP Gaussian-process hyper-parameters $\Lambda$. We find $k_0=3.13\times 10^{-39}, w=\unit[3.0]{cycles}, l=\unit[2.0]{cycles}$ and $f_0= 0.0633 M^{-1}$. 
Using the MAP estimate hyper-parameters, we estimate the mean deviated strain $\tilde{s}$ for each event using Eq.~\ref{eq:predicted_mean}, as well as its covariance $\Sigma$ using Eq.~\ref{eq:predicted_covariance}.

We use $\Delta M = 10M_\odot$.
With $28$ events, spanning a total mass range of $\sim 70 M_\odot$, the average spacing between events is $\sim 2.5 M_\odot$. Choosing  $\Delta M =10M_\odot$ ensures nearby events can be combined coherently. We show results for different choices of $\Delta M$ in Appendix \ref{appendix:other_delta_M}.

\begin{figure*}
    \centering
    \includegraphics[width=18cm]{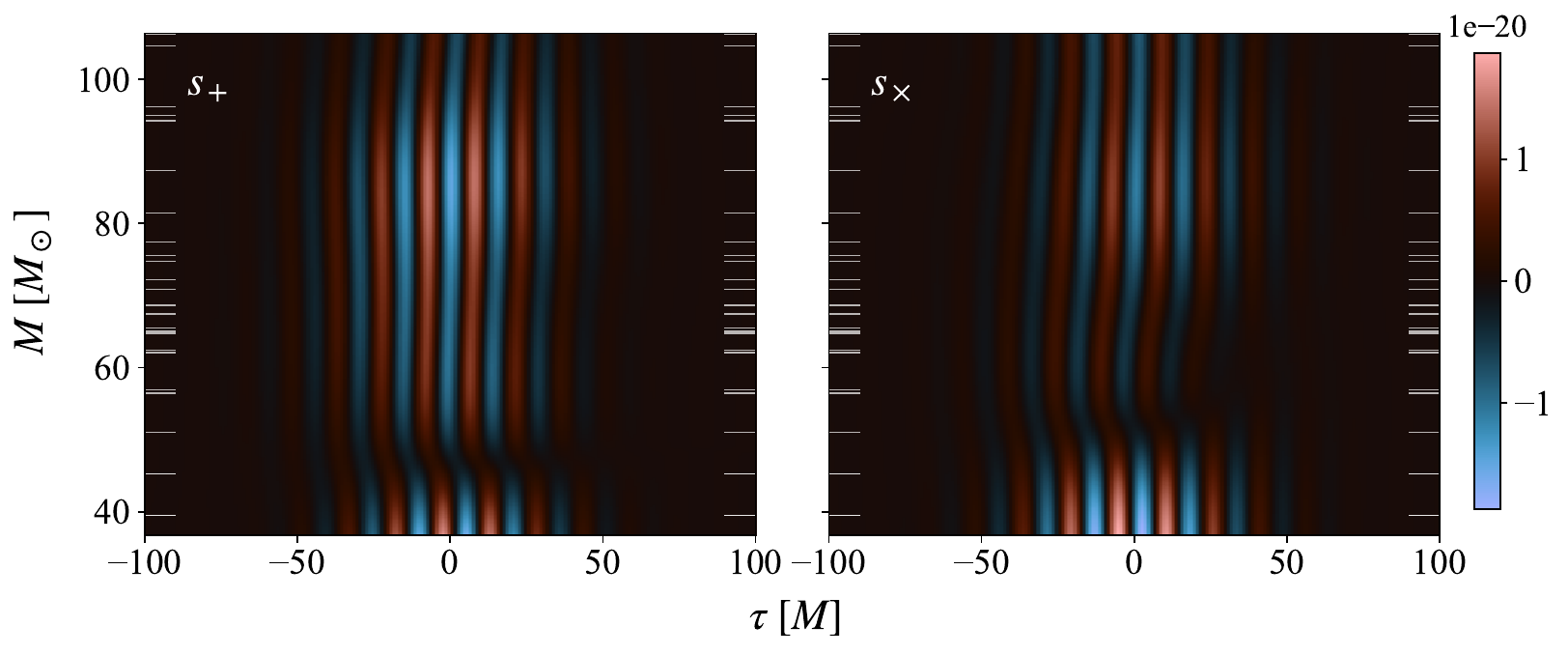}
    \caption{Heatmap of deviated strain using information from coherently combining $28$ GWTC-5 events, for the + (left) and $\times$ (right) polarisations. Horizontal dashed lines indicate the positions of events. While a coherent signal seems mildly supported by the data, it is not statistically significant. }
    \label{fig:gwtc_map}
\end{figure*}

We show the resulting maps of $\hat{s}$ for each polarisation in Fig.~\ref{fig:gwtc_map}.
We calculate $\lambda_\text{pop}$: the likelihood ratio comparing the hypothesis that $\hat{s}>0$ to the hypothesis that $\hat{s}=0$ and obtain $\ln\widehat\lambda_\text{pop} = 0.46$ ($p=0.05$), which means the data are consistent with general relativity. 

\section{Assumptions and approximations}\label{section:discussion}
In this Section, we discuss the many assumptions and approximations employed in our analysis.
We see this work as providing foundational ideas upon which a more complicated framework could be built.
By listing our assumptions systematically, our aim is to highlight what we regard as a minimalist phenomenology for deviations from general relativity.
Our approximations are another matter.
Some of our approximations are entirely defensible.
However, in order to deal with the computational challenges brought on by our model, we employ here some approximations that we regard as---at best---inelegant.
We list all of our approximations systematically to highlight what we think can be improved in future implementations.

\subsection{Assumptions}\label{subsection:assumptions}

We divide our assumptions into two categories.
\textit{Standard} assumptions are those that are commonly used in LVK analyses. We contrast these with \textit{novel} assumptions.

\subsubsection{Standard assumptions}
\begin{itemize}
    \item \textbf{Deviated strain is $+,\times$-polarized.} 
    General metric theories of gravity permit up to six gravitational-wave polarisation modes. 
    \item \textbf{The deviated strain is not dispersive.} 
    In other words, propagation speed is frequency-independent.
\end{itemize}

\subsubsection{Novel assumptions}
\begin{itemize}
    \item \textbf{Events with identical binary parameters exhibit identical deviated strain.}
    This implies our model is deterministic and that there are no ``hidden variables'' that can affect the deviated strain.
    \item \textbf{The deviated strain map is slowly varying.}
    If this assumption were not true, then two very similar events with slightly different binary parameters, could have completely different deviated strains.
    \item \textbf{In practice, the deviated strain depends only on the total mass.}
    This assumption together with the previous two imply that deviations from GR can be represented by something like Fig.~\ref{fig:2d_map_example}.
    In principle, one could allow the deviated strain map to depend on additional variables.
    \item \textbf{The deviated strain is most pronounced near the merger and oscillates with a frequency that is not dramatically different from the ringdown frequency.}
    We enforce this assumption through our choice of Gaussian process kernel.
    \item \textbf{Deviated strain is carried primarily by the $(2,2)$ mode}. 
    Without this assumption it is difficult to compare identical events viewed from different angles.
    In principle, one could generalize the analysis to construct deviated strain maps for multiple modes.
\end{itemize}

\subsection{Approximations}\label{subsection:approximations}
We distinguish between approximations that we regard as suitable (marked with $\checkmark$) and those that we would prefer to not have to make in future analyses (marked with \ding{55}).
\begin{itemize}

\item[\checkmark] \textbf{Waveform approximants are adequately specified.} 
We assume waveform systematics are negligible.
If this assumption does not hold, we expect the deviated strain map to show statistically significant structure.
However, since GR is scale invariant, deviated strain should not vary with mass, thereby providing a powerful tool for discriminating waveform error from deviations from GR.

\item[\checkmark] \textbf{The deviated strain waveforms are mostly orthogonal to the GR signal.}
This assumption is implicit in our analysis because we first fit the GR signal and then fit the residuals.
However, if the two signals are similar, the deviated strain can be absorbed into the GR fit.
That said, as best we can tell, the only thing we can measure is the deviated strain orthogonal to the binary waveform.

\item[\ding{55}] \textbf{We ignore waveform subtraction uncertainty.} 
Since we do not measure the GR waveform perfectly, we should marginalize over the uncertain binary parameters.
We are not able to do this in our frequentist formulation.
While our deviated strain map preserves qualitative features of the true map, it does not provide a statistically rigorous characterization of the true map.

\item[\ding{55}] \textbf{We ignore uncertainty in the Gaussian-process hyper-parameters.}
Our Gaussian process is characterized by hyper-parameters that determine the morphology of the deviated strain. 
These hyper-parameters are constrained, but not determined, by the data.
By ignoring this uncertainty, our deviated strain map loses an additional layer of statistical rigour.

\item[\ding{55}] \textbf{We ignore total-mass uncertainty.}
The deviated-strain point estimates for each event are added to our map based on their total mass.
However, this quantity is not precisely known.
By ignoring this uncertainty, features in the true deviated strain map will tend to be somewhat smeared out in our reconstructions.
This smearing is largely ameliorated by our SNR cut, but as a result, we are forced to exclude some data.

\item[\ding{55}] \textbf{We ignore uncertainty in viewing angle.}
A key step in our calculation is to transform the signal measured in the detectors to the fiducial reference frame.
This transformation relies on measurements of various extrinsic binary parameters.
By ignoring the uncertainty on these parameters, our deviated strain map loses yet another layer of statistical rigour.
\end{itemize}

Despite the fact that so many of these approximations can be improved, we are heartened by Fig.~\ref{fig:2d_gp_injection}, which shows that we are able to recover something qualitatively similar to the injected strain map in spite of several approximations that we would rather not have to make.

\section{Conclusions}\label{subsection:conclusions}
We introduce a framework for searching for deviations from general relativity in a catalog of gravitational-wave events.
We introduce the idea of a \textit{deviated strain map}, which uses a fiducial reference frame to connect events with similar binary parameters.
We also introduce the idea that the deviated strain map varies slowly with total mass.
We point out mass-dependence can be used to distinguish waveform misspecification (which should be mass-independent) from new physics.
We suggest this is a promising starting point for building a phenomenology of deviations from GR, which can be used to rigorously distinguish new physics from scale-invariant waveform misspecification.
Using this framework, we analyse $28$ binary-black hole mergers from the fifth gravitational wave transient catalog, and find no evidence for deviations from GR ($p=0.05$).

We view the analysis carried out in this manuscript as a proof of concept.
We suggest that a straightforward Bayesian implementation of our model is computationally intractable.
Our frequentist implementation is adequate to show the potential of this framework, but within that framework, we are forced to make a number of approximations that we would rather remove in the next iteration of this work.
Here we sketch out a possible path forward.

The first step in our new design is dimensional reduction: to find some set of parameters $\Upsilon$, which can be used to efficiently characterize the deviated strain maps drawn from our generative model.
For example, it may turn out that, say, 25 principal-component maps can be used to represent the vast majority of maps.
In this case, $\Upsilon$ would correspond to the set of 25 principle-component coefficients.
The key point of this step is to find a way to represent the deviated strain map with some set of parameters that is very small compared to the ${\cal O}(10^{10})$ pixels that make up the map.

Having identified this compact set of parameters $\Upsilon$, the next step would be to find the prior $\pi(\Upsilon)$ implied by our generative model.
Since samples of $\pi(\Upsilon)$ can be generated relatively rapidly ($\lesssim\unit[1]{s}$ per core), the distribution $\pi(\Upsilon)$ can probably be fit empirically, e.g., using a normalizing flow.

The next step is to analyze each event individually with just the binary hypothesis.
This yields samples for the binary posterior
\begin{align}\label{eq:fiducial}
    p_{\o}(\theta_i | d_i) = 
    \frac{1}{{\cal Z}(d_i)} 
    {\cal L}_{\o}(d_i | \theta_i) \pi(\theta_i) ,
\end{align}
which can be subsequently used for importance sampling.\footnote{In Eq.~\ref{eq:fiducial}, the subscript $i$ refers to event $i$ and the subscript $\o$ is used to denote distributions for imperfect hypotheses, used for importance sampling.}
The likelihood for the map parameters $\Upsilon$ can then be written like so:
\begin{align}
    {\cal L}(\vec{d} | \Upsilon) =
    \prod_i^N \frac{1}{n}
    \sum_k^n \frac{
    {\cal L}(d_i | \theta_i^k, \Upsilon) \pi(\theta_i^k)}
    {{\cal L}_{\o}(d_i | \theta_i^k)} .
\end{align}
The sum over $k$ is a sum over the binary posterior samples.
By writing the integral like this, we assume that the posterior for the binary parameters does not change too much due to the presence of deviations from GR.
However, one can at least know that whether this is a suitable approximation by checking the number of effective samples in the sum; see, e.g., \cite{hom}.
(One can even tweak the proposal distribution to improve convergence.)

This approach seems promising.
Thanks to the dimensional reduction, it does not require inversion of enormous matrices.
It completely removes all of the approximations that we introduced in this paper by marginalizing over uncertainty in the binary parameters.
It would still be a complicated workflow with a number of intricate steps.

\section*{Acknowledgements}
This is LIGO document \#P2600561.
We acknowledge support from the Australian Research
Council (ARC) Centres of Excellence CE170100004 and CE230100016, as well as ARC
LE210100002, and ARC DP230103088. L.P. receives support from the Australian Government Research
Training Program. 
This material is based upon work supported by NSF’s LIGO Laboratory
which is a major facility fully funded by the National
Science Foundation. The authors are grateful for computational resources provided by the LIGO Laboratory
and supported by National Science Foundation Grants
PHY-0757058 and PHY-0823459.

This research has made use of data or software obtained from the Gravitational Wave Open Science Center (gw-openscience.org), a service of LIGO Laboratory,
the LIGO Scientific Collaboration, the Virgo Collaboration, and KAGRA. LIGO Laboratory and Advanced
LIGO are funded by the United States National Science Foundation (NSF) as well as the Science and Technology Facilities Council (STFC) of the United Kingdom, the Max-Planck-Society (MPS), and the State of
Niedersachsen/Germany for support of the construction
of Advanced LIGO and construction and operation of
the GEO600 detector. Additional support for Advanced
LIGO was provided by the Australian Research Council.
Virgo is funded, through the European Gravitational
Observatory (EGO), by the French Centre National
de Recherche Scientifique (CNRS), the Italian Istituto
Nazionale di Fisica Nucleare (INFN) and the Dutch
Nikhef, with contributions by institutions from Belgium,
Germany, Greece, Hungary, Ireland, Japan, Monaco,
Poland, Portugal, Spain. The construction and operation of KAGRA are funded by Ministry of Education,
Culture, Sports, Science and Technology (MEXT), and
Japan Society for the Promotion of Science (JSPS), National Research Foundation (NRF) and Ministry of Science and ICT (MSIT) in Korea, Academia Sinica (AS)
and the Ministry of Science and Technology (MoST) in
Taiwan.

\section*{Data Availability}

The data underlying this article are publicly available at \url{https://www.gwosc.org}.
\bibliographystyle{aasjournal} 
\bibliography{refs}
\appendix

\section{Choosing the test grid}\label{appendix:choosing_tau}
The chosen natural-time coordinate $\tau$ should comfortably contain the deviated strain for all possible hyper-parameter values allowed by our priors (see Section \ref{section:injection_study}). It should also avoid undersampling below the Nyquist limit, and should not oversample too much, to reduce computational cost.

The envelope kernel term $\exp(-f_0^2 (\tau_a^2 + \tau_b^2)/2w^2)$, for one time coordinate $\tau_a = \tau_b$, becomes $\exp(-f_0^2 \tau_a^2/w^2)$, which is Gaussian in $\tau_a$ with width $\sigma_\text{env} = w/(f_0\sqrt{2})$. With $w=3.0$, the widest possible envelope under our priors occurs at the lower bound of $f_0$ ($w=3.0$, $f_0=0.046$), giving $\sigma_\text{env} \approx 46.1$ natural-time units. A $\tau$ range of $[-400, 400]$ therefore captures the worst case envelope to $\approx 9\sigma$, which means almost no deviated-strain power is truncated at the edges of $\tau$ even for the most extreme possible hyper-parameters.

The oscillatory term $\cos(2\pi f_0|\tau_a-\tau_b|)$ must be resolved (without aliasing) up to the largest natural frequency our prior allows, $f_0=0.108$, requiring a grid spacing $\Delta\tau < 1/(2f_0) \approx 4.63$. We choose a grid spacing of $\Delta\tau = 1$, which oversamples this limit by a factor of $4.63$, allowing for a wide margin against aliasing without adopting a finer grid than necessary.

\section{Events analysed}\label{appendix:events_analysed}
In Table \ref{table:event_choices}, we list the events passing our selection criteria (Subsection \ref{subsection:event_selection_criteria}) that we analyse in Section \ref{section:GWTC-5}.

\begin{deluxetable}{lcc|lcc}\label{table:event_choices}
\tablecaption{The 28 GWTC-5 events analysed in Section~\ref{section:GWTC-5}, passing the selection criteria of Subsection~\ref{subsection:event_selection_criteria}, ordered by source-frame total mass.\label{tab:events}}
\tablehead{
\colhead{Event} & \colhead{$M$ [$M_\odot$]} & \colhead{Network SNR} & \colhead{Event} & \colhead{$M$ [$M_\odot$]} & \colhead{Network SNR}
}
\startdata
GW240930\_035959 & 36.82 & 17.68 & GW231206\_233901 & 67.43 & 21.19 \\
GW230927\_153832 & 39.46 & 19.06 & GW240920\_124024 & 68.65 & 37.24 \\
GW190412\_053044 & 39.55 & 18.69 & GW240919\_061559 & 70.91 & 16.14 \\
GW190408\_181802 & 45.24 & 16.70 & GW240621\_195059 & 72.18 & 26.74 \\
GW241006\_015333 & 45.25 & 16.84 & GW240511\_031507 & 74.73 & 17.44 \\
GW230919\_215712 & 51.06 & 15.58 & GW231226\_101520 & 75.52 & 34.16 \\
GW170814\_103043 & 56.46 & 16.99 & GW200224\_222234 & 77.39 & 18.72 \\
GW190828\_063405 & 56.92 & 15.61 & GW190521\_074359 & 81.46 & 25.67 \\
GW240615\_113620 & 62.04 & 21.84 & GW241127\_061008 & 87.37 & 31.70 \\
GW150914\_095045 & 62.12 & 23.86 & GW240705\_053215 & 94.23 & 15.25 \\
GW200129\_065458 & 62.37 & 26.79 & GW240514\_121713 & 95.02 & 16.02 \\
GW230628\_231200 & 64.76 & 15.43 & GW230914\_111401 & 96.20 & 15.34 \\
GW200311\_115853 & 65.02 & 16.99 & GW241225\_082815 & 104.66 & 18.00 \\
GW250114\_082203 & 65.50 & 77.29 & GW190706\_222641 & 106.28 & 15.05 \\
\enddata
\end{deluxetable}

\section{Simulated-data generation}\label{appendix:data_generation}
As our framework describes an \textit{estimator} for deviated strain, and does not describe a generative model, it is not possible to simulate a deviation from GR directly.
Therefore, we simulate each polarisation of the deviated strain as draws from a two-dimensional Gaussian process, described by the kernel
\begin{align}\label{eq:K_m}
    K_{ab} \equiv & k(\tau_a, \tau_b, M_a, M_b) \nonumber\\
    = &
    k_0 \exp{\left(-f_0^2\frac{\tau_a^2+\tau_b^2}{2 w^2}\right)}
    \exp{\left(-f_0^2\frac{|\tau_a - \tau_b|^2}{2l_t^2}\right)}
    \cos\left(2\pi f_0 |\tau_a - \tau_b|\right)
    \exp{\left(-\frac{|M_a - M_b|^2}{2l_M^2}\right)},
\end{align}

where $(\tau_a, \tau_b)$ are natural time points in units of $M$, $(M_a, M_b)$ are total mass points in units of $M_\odot$, $k_0$ describes the standard deviation of the deviated strain at the merger time, $w$ and $l_t$ are length scales in units of cycles, $f_0$ is a characteristic oscillation frequency, and $l_M$ is a length scale in the total-mass direction, and can be loosely identified with the $\Delta M$ of our model.

We draw independent realisations for each deviated-strain polarisation from the kernel in Eq.~\ref{eq:K_m} on an evenly spaced $\tau$ grid of $1024$ points in the range $(-512 M, 512 M)$. We then interpolate it onto our detector-frame time points $t$ of $(\unit[-1]{s},\unit[1]{s})$, with a sampling frequency of $\unit[1024]{Hz}$. To transform to the detector frame, we use the binary parameters for the same $28$ events from GWTC-5. We perform the inverse of the operations in Eq.~\ref{eq:transformation} to obtain detector-frame data, with the appropriate inclination and coalescence-phase transformations, scaled by the luminosity distance of each event, and interpolated onto uniformly spaced physical frequency points, in units of $\text{Hz}$. 

\section{Joint-likelihood gradients}\label{appendix:gradients}
To find the best-fit hyper-parameters for the posterior in Eq.~\ref{eq:joint_posterior}, we optimise for the maximum-a-posteriori estimate using the \textsc{L-BFGS-B} algorithm. We work with the negative logarithm of the posterior. The \textsc{L-BFGS-B} algorithm is faster and more numerically stable if it is provided with analytic gradients (with respect to the hyper-parameters). We derive the gradient for individual events first, then the multi-event case follows by summation. Throughout, we work with logarithms of the hyper-parameters. 

First, we calculate the derivatives of the per-event likelihoods --- we construct the posterior derivatives at the end.
For a single event, we model the residual data  $r$ as a Gaussian process with total covariance
\begin{align}
    C(\Lambda) = S(\Lambda) + N,
\end{align}
where $N$ is the noise covariance from Eq.~\ref{eq:N}, and $S$ is the deviated-strain covariance, built from the shared hyper-parameters $\Lambda$. Using the same Whittle-likelihood form as Eq.~\ref{eq:single_event_likelihood},
\begin{align}
    \ln\mathcal{L}(r \mid \Lambda) = -n\ln\pi - \ln\det C - r^\dagger C^{-1} r.
\end{align}

The standard identities for the derivative of a Gaussian log-likelihood with respect to covariance hyper-parameters $\Lambda$ \citep[see, e.g.][]{rasmussen_gaussian_2008} are
\begin{align}
    \frac{\partial}{\partial\Lambda}\ln\det C &= \mathrm{tr}\!\left(C^{-1}\frac{\partial C}{\partial\Lambda}\right), \nonumber \\
    \frac{\partial}{\partial\Lambda}\left(r^\dagger C^{-1} r\right) &= -(C^{-1}r)^\dagger \frac{\partial C}{\partial\Lambda}(C^{-1}r),
\end{align}
so that
\begin{equation}\label{eq:grad_general}
    \frac{\partial \ln\mathcal{L}(r \mid \Lambda)}{\partial\Lambda} = -\mathrm{tr}\!\left(C^{-1}\frac{\partial C}{\partial\Lambda}\right) + (C^{-1}r)^\dagger \frac{\partial C}{\partial\Lambda}(C^{-1}r).
\end{equation}

Here, $C$ is formed using a signal covariance matrix $S(\Lambda)$ and noise matrix $N$, of which only $S(\Lambda)$ depends on $\Lambda$. We obtain $S$ from the natural-time-domain kernel matrix $K_\tau(\Lambda)$ (defined in Eq.~\ref{eq:Ktau}) using Eq.~\ref{eq:Sij}. Because the operations applied to $K_\tau$ are all linear, we simply write
\begin{equation}\label{eq:linear_map}
    \frac{\partial C}{\partial\Lambda} = \mathcal{F}^\dagger\frac{\partial K_\tau}{\partial\Lambda}\mathcal{F},
\end{equation}
where $\mathcal{F}$ is the collection of linear operations performed in Eq.~\ref{eq:Sij}.
This means we can differentiate the natural-time kernel analytically, then apply the same operator $\mathcal{F}$ to each derivative.

The derivatives of $C$ with respect to the hyper-parameters are
\begin{align}\label{eq:single_event_derivatives}
    \frac{\partial C}{\partial \ln k_0} = & \ \mathcal{F}^\dagger K_\tau \mathcal{F} , \nonumber \\
    \frac{\partial C}{\partial \ln w} = & \ \mathcal{F}^\dagger K_\tau \frac{f_0^2(\tau_a^2 + \tau_b^2)}{w^2} \mathcal{F}, \nonumber \\
    \frac{\partial C}{\partial \ln l} = & \ \mathcal{F}^\dagger K_\tau \frac{f_0^2 (\tau_a-\tau_b)^2}{l^2}\mathcal{F}, \nonumber \\
    \frac{\partial C}{\partial \ln f_0} = & \ -\mathcal{F}^\dagger [ K_\tau f_0^2 \left(\frac{(\tau_a^2 + \tau_b^2)}{w^2} + \frac{(\tau_a-\tau_b)^2}{l^2}\right) \nonumber \\ & + k_0 \exp\left(-f_0^2\frac{(\tau_a^2 + \tau_b^2)}{2w^2}\right)\exp\left(-f_0^2\frac{(\tau_a-\tau_b)^2}{2l^2}\right)\sin(2\pi f_0 |\tau_a-\tau_b|) 2\pi f_0 |\tau_a - \tau_b|]\mathcal{F} ,
\end{align}
where we have included derivatives with respect to $w$ and $l$ for completeness, though in this work, we fix these to values of $3.0$ and $2.0$ respectively.

Writing $d \equiv \{d_i\}_{i=1}^{Q}$ for the residual data of all events, Bayes' theorem gives
\begin{equation}\label{eq:joint_posterior_gradients}
    -\ln p(\Lambda \mid d) = \sum_{i=1}^{Q} \left[-\ln\mathcal{L}_i(\Lambda_i)\right] -\ln\pi(\Lambda) + \ln{Z},
    \qquad \Lambda_i \equiv \Lambda - (0,0,0,\ln\beta_i),
\end{equation}
where we have accounted for the fact that the only difference in $\Lambda$ between events is the characteristic frequency, by subtracting a factor $\ln \beta_i$ (see Eq.~\ref{eq:lambda_k}), $\pi(\Lambda)$ is the hyper-parameter prior of Section~\ref{section:injection_study} and $Z$ is the evidence, which we can ignore. The posterior gradients are then

\begin{align}\label{eq:joint_gradient}
    -\frac{\partial \ln p(\Lambda\mid d)}{\partial \Lambda}
    = \sum_{i=1}^{Q} \left.\frac{\partial\left[-\ln\mathcal{L}_i\right]}{\partial \Lambda}\right|_{\Lambda_i}
    - \frac{\partial \ln\pi(\Lambda)}{\partial \Lambda} .
\end{align}

To summarise, the posterior gradients are obtained by first taking the hyper-parameter derivative for each individual event, adding these contributions for all events, and subtracting the gradient of the prior.

\section{Detection statistic}\label{appendix:detection_statistic}
We calculate a detection statistic for the hypothesis that, for each event, the coherently combined deviated strain is greater than zero. The deviated strain for event $k$ predicted by the other events is (see Eq.~\ref{eq:stacking_for_map})
\begin{align}\label{eq:prediction_mean}
    {\hat{s}}_i = \Pi_i \sum_{j<i}^{Q} w_{ij}(\Delta M) \Sigma^{-1}_{j} {\tilde{s}}_j ,
\end{align}
with associated covariance describing the uncertainty in this prediction
\begin{align}\label{eq:prediction_covariance}
    \Pi_i = \left( \sum_{j<i}^{Q} w_{ij}(\Delta M) {\tilde{\Gamma}}_j \right)^{-1} .
\end{align}
Here, events are ordered by their maximum-likelihood source-frame total mass. The sum over $j<i$ ensures event $i$ does not enter its own prediction. 

To see why, consider the following. The total evidence ${\cal Z}(d_{1:Q})$ can be factored by the
probability chain rule. Take the last event, conditioned on all the
preceding ones,
\begin{align}
    {\cal Z}(d_{1:Q}) = {\cal Z}(d_Q \mid d_{1:Q-1})\,{\cal Z}(d_{1:Q-1}) .
\end{align}
The remaining term ${\cal Z}(d_{1:Q-1})$ factors the same way,
\begin{align}
    {\cal Z}(d_{1:Q-1}) = {\cal Z}(d_{Q-1} \mid d_{1:Q-2})\,{\cal Z}(d_{1:Q-2}) .
\end{align}
Iterating to the first event, by induction, the total evidence is
\begin{align}
    \ln {\cal Z}(d_{1:Q}) = \sum_{i=1}^{Q} \ln {\cal Z}(d_i \mid d_{1:i-1}) ,
\end{align}
where the first event is unconditioned, but contributes to the evidence for all other events.

We compare two hypotheses. Under the null, $\mathcal{H}_0$, there is no deviated strain signal and the data is explained by just the noise,
\begin{align}\label{eq:null_likelihood}
    \mathcal{L}(r_i \mid \mathcal{H}_0)
    = \frac{1}{\pi^n \det N_{i}}
      \exp\left( -r_i^\dagger N_{i}^{-1}
      r_i \right) ,
\end{align}
where $r_i$ and $N_i$ are interpolated onto the $\tau$ grid for the $i$th event.

Under the alternative, $\mathcal{H}_1$, a non-zero deviated strain signal is present and is
predicted by the other events,
\begin{align}\label{eq:alt_likelihood}
    \mathcal{L}(r_i \mid \mathcal{H}_1)
    = \frac{1}{\pi^n \det(N_{i} + \Pi_i)}
    \exp\left( -(r_i - {\hat{s}}_i)^\dagger
    (N_{i} + \Pi_i)^{-1}
    (r_i - {\hat{s}}_i) \right) .
\end{align}

The per-event likelihood ratio is then
\begin{align}\label{eq:per_event_detection_statistic}
    \lambda_i =
      \frac{\mathcal{L}(r_i \mid \mathcal{H}_1)}
           {\mathcal{L}(r_i \mid \mathcal{H}_0)} .
\end{align}
The total significance for the dataset is just the product over events,
\begin{align}\label{eq:population_detection_statistic}
    \lambda_\text{pop} = \prod_{i=1}^{K} \lambda_i .
\end{align}

\section{Background Study}\label{appendix:background_study}
In the absence of an analytic form for the detection--statistic background distribution, we calibrate the measured detection statistic $\widehat{\lambda}_\text{pop}$ against an empirical background distribution, $\pi(\lambda_\text{pop}\mid \mathcal{H}_0)$.

To form a statistically significant background, we adopt a similar procedure as Subsection {\ref{section:injection_study}}, where we aim produce simulated data similar to the real data we analyse. We form data following the exact procedure as in Subsection {\ref{section:injection_study}} for $N=28$ events, except we do not inject a deviated strain signal. We perform the same analysis procedure as before. 

We calculate the population detection statistic $\lambda_\text{pop}$ (see Appendix \ref{appendix:detection_statistic}).  We repeat this procedure $97$ times (i.e. we analyse $97$ simulated catalogs of $28$ events containing only noise). 

\begin{figure}
    \centering
    \includegraphics[width=0.5\linewidth]{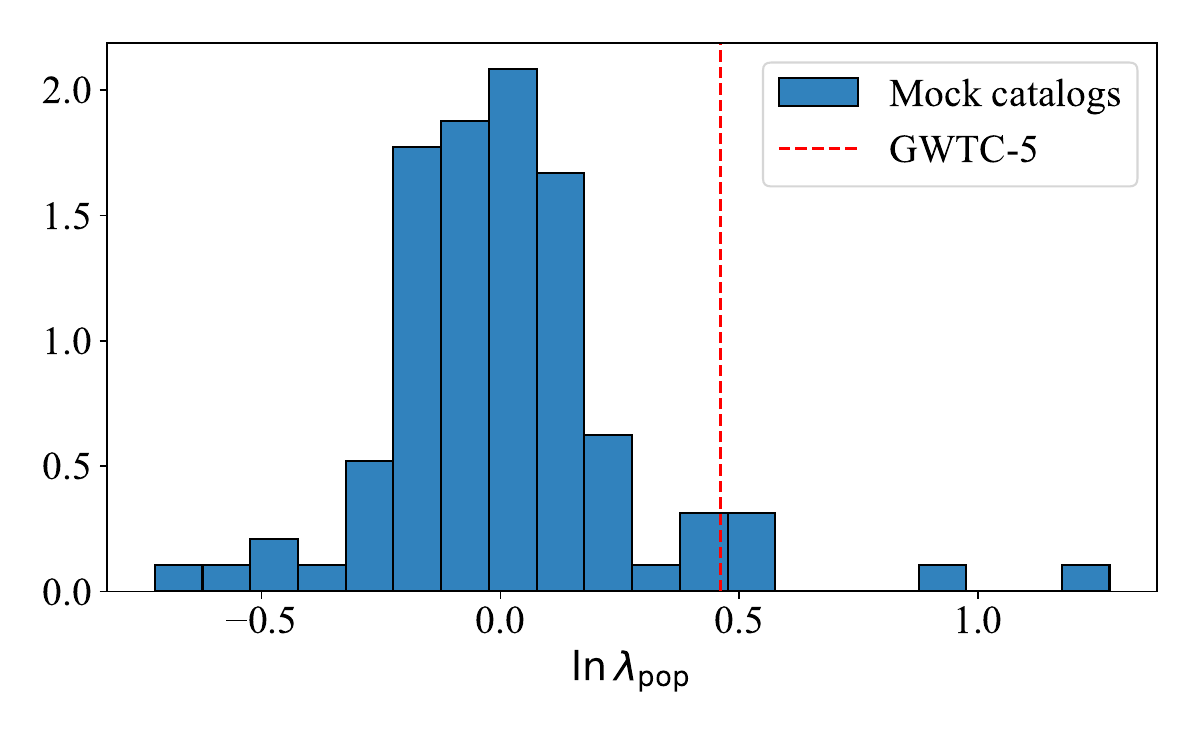}
    \caption{Comparison between an empirical distribution for the population-level detection statistic $\pi(\ln\lambda_\text{pop})$ from $97$ off-source data segments and the $\ln\lambda_\text{pop}$ calculated for GWTC-5 events. Not shown is a single off-source catalog with $\ln\lambda_\text{pop}\approx-6$. Though the result for GWTC-5 lies in the tail of the distribution for mock catalogs, it is still consistent, with $p=0.05$.}
    \label{fig:background_study}
\end{figure}

We calculate an empirical distribution for $\lambda_\text{pop}$, $\pi(\lambda \mid \mathcal{H}_0)$. We show this distribution in Figure \ref{fig:background_study} and also show the measured $\widehat\lambda_\text{pop}$ for GWTC-5. From this distribution, we can calculate a $p$-value for the measured $\widehat\lambda_\text{pop}$ by counting the number of catalogs in the background with $\lambda_\text{pop}>\widehat\lambda_\text{pop}$,
\begin{equation}\label{eq:p-value}
      p = \int_{\widehat\lambda_\text{pop}}^{\infty} \pi(\lambda_\text{pop} \mid \mathcal{H}_0) \, d\lambda_\text{pop} .
\end{equation}
We find $p=0.05$ for GWTC-5, suggesting no evidence for deviations from GR.

\section{Other choices for $\Delta M$}\label{appendix:other_delta_M}
\begin{figure}[]
    \centering
    \includegraphics[width=18cm]{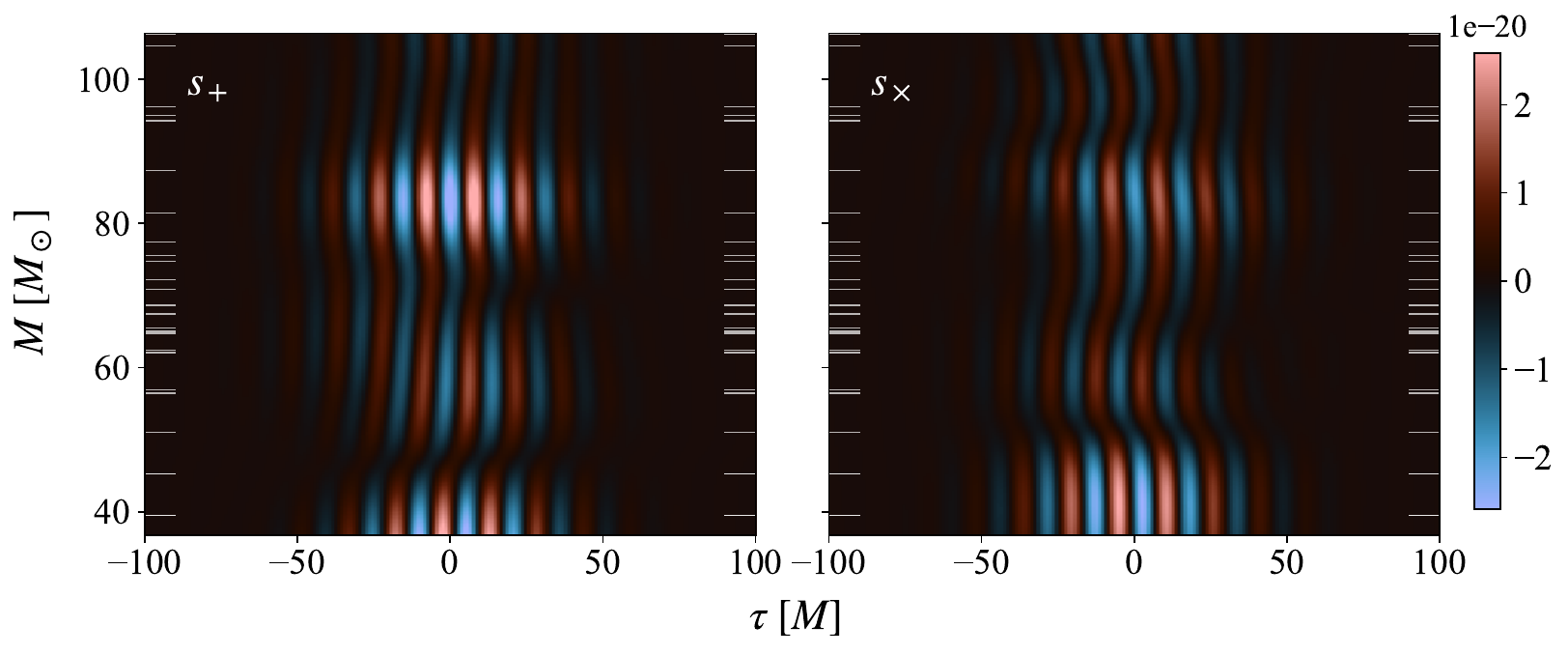}
    \caption{As in Figure \ref{fig:gwtc_map}, but for $\Delta M= 5$}
    \label{fig:gwtc_map_deltaM_5}
\end{figure}

\begin{figure}[]
    \centering
    \includegraphics[width=18cm]{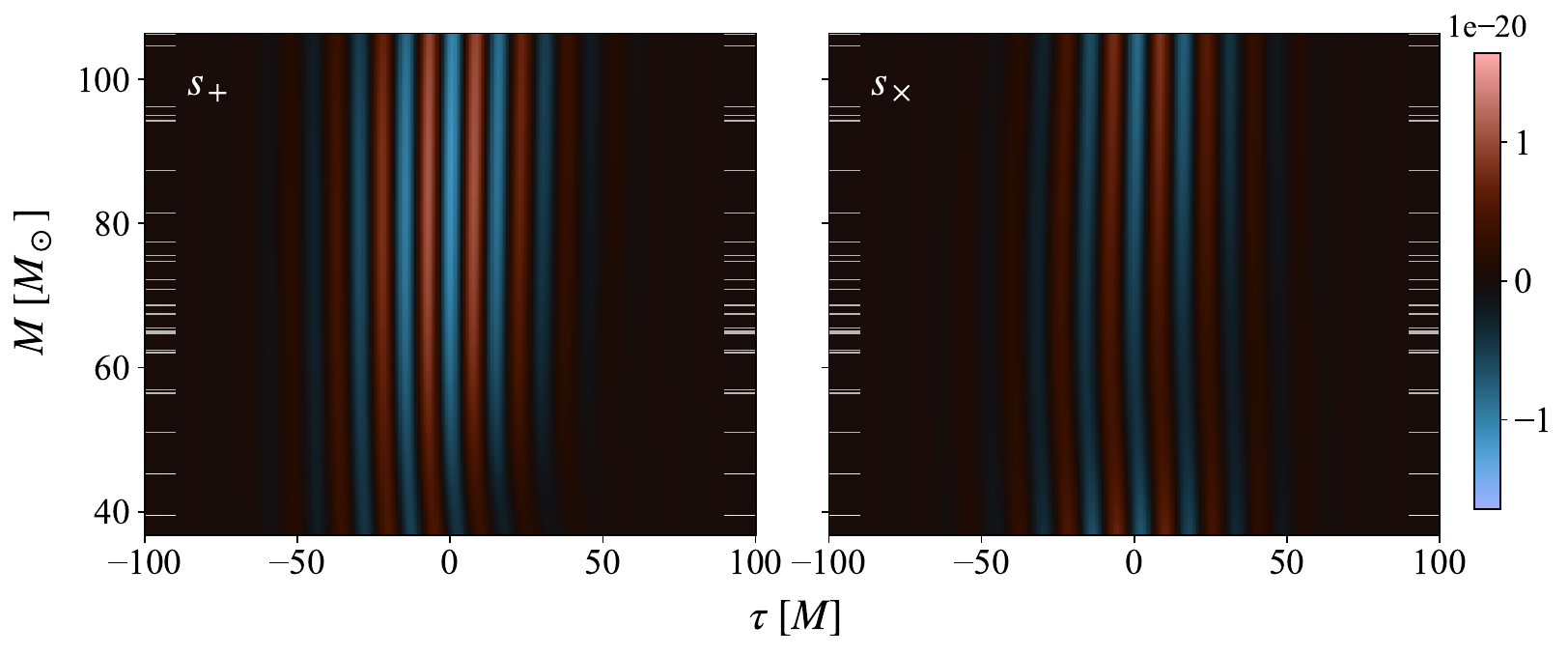}
    \caption{As in Figure \ref{fig:gwtc_map}, but for $\Delta M= 20$}
    \label{fig:gwtc_map_deltaM_20}
\end{figure}

In this work, we choose a coherence length scale total mass $\Delta M= 10$. Other choices are possible, but we make this choice as our event selection cuts imply our events can only be meaningfully localised in a range of $\approx 10 M_\odot$ (see Subsection \ref{subsection:event_selection_criteria}). We show other possible choices for our analysis of a population of $28$ GWTC-5 events in Figures \ref{fig:gwtc_map_deltaM_5} and \ref{fig:gwtc_map_deltaM_20}. The signal amplitude remains low, indicating no deviations from GR, though we do not calculate detection statistics for these analyses.

\end{document}